\documentclass[sigconf]{acmart}

\usepackage{multirow}   
\usepackage{makecell}

\usepackage{graphicx}
\usepackage{textcomp}
\usepackage{hyperref}
\usepackage{float}
\usepackage{balance}
\usepackage{subcaption} 
\usepackage{soul}
\usepackage{multirow}

\AtBeginDocument{%
  }

\setcopyright{cc}
\setcctype[4.0]{by}
\copyrightyear{2026}
\acmYear{2026}

\acmDOI{10.1145/3772318.3790336}

\acmConference[CHI '26]
{2026 CHI Conference on Human Factors in Computing Systems}
{April 13--17, 2026}
{Barcelona, Spain}

\acmISBN{979-8-4007-2278-3/2026/04}

\begin{document}




\title{Oops, I Did It Again (But I Know It):
Robot Failure Consistency and Awareness
in Human-Robot Collaboration}


\author{Ramtin Tabatabaei}
\affiliation{%
  \institution{The University of Melbourne}
  \city{Melbourne}
  \country{Australia}}
\email{stabatabaeim@student.unimelb.edu.au}

\author{Vassilis Kostakos}
\affiliation{%
  \institution{The University of Melbourne}
  \city{Melbourne}
  \country{Australia}}
\email{vassilis.kostakos@unimelb.edu.au}

\author{Wafa Johal}
\affiliation{%
  \institution{The University of Melbourne}
  \city{Melbourne}
  \country{Australia}}
\email{wafa.johal@unimelb.edu.au}

\renewcommand{\shortauthors}{ }

\begin{abstract}

In human–robot collaboration, repeated failures are inevitable and can undermine trust and perceptions of robot intelligence. While some failures severely disrupt tasks and others are relatively benign, their cumulative impact on trust is not clearly understood. We investigated whether users perceive repeated failures of the same type differently from varied failures, and how robot awareness of its own failures affects these perceptions. In a collaborative physical task with 54 participants, we manipulated failure sequence (homogeneous vs. heterogeneous) and awareness (none, partial, full). Results show that trust and perceived intelligence were influenced by both current and prior failures, with homogeneous sequences leading to smaller reductions in these evaluations compared to heterogeneous ones. Robots displaying awareness, whether partial or full, were consistently rated higher than unaware robots, particularly for grasping and planning failures. Our findings provide a deeper understanding of how failure type, sequence, and robot awareness shape users'  perceptions of collaborative robots.

\end{abstract}




\keywords{Failure sequence, Trust repair, Human-Robot Collaboration}


\maketitle

\section{Introduction}

In human–robot interaction (HRI), failures are unavoidable, particularly when robots operate in uncertain and dynamic real-world environments \cite{benali_dilemma_2025, jung_robots_2018, schombs2024b}. These failures can occur not only once but also repeatedly throughout an interaction. While single failures have been widely studied in HRI \cite{salem_evaluating_2015, zhang_effects_2023}, repeated failures introduce additional complexities. They have been shown to produce cumulative erosion of trust and shift expectations in human–human contexts \cite{lewicki_trust_2017, tomlinson_road_2004}, and emerging work suggests similar dynamics may also occur in HRI \cite{esterwood_repairing_2025}. Crucially, not all failures are perceived equally \cite{gideoni_is_2024}. The perceived severity of failures within a sequence can shape user evaluations of a robot differently than a single failure in isolation, because sequential failures influence how user expectations develop over time. This pattern aligns with Expectancy Violation Theory (EVT) \cite{burgoon_interpersonal_1993, burgoon_nonverbal_1988}, which predicts stronger negative reactions when events deviate from emerging expectations.

To better interpret how users respond to repeated failures, it is also important to consider where these failures come from and how they manifest. Such failures may arise from a single factor, such as limited sensory perception or hardware constraints, or from their combination with the complexities of dynamic environments \cite{desai_creating_2009, honig_understanding_2018}. These factors not only trigger failures but also influence their manifestation patterns. A sequence of robot failures can be categorised into two broad types: homogeneous failures, where the exact same failure recurs under the same operating environment; and heterogeneous failures, where different types of failures arise within the same environment. For example, a homogeneous failure may involve repeated joint fatigue due to a robot performing a repetitive task \cite{lamon_towards_2018}, or consistent localisation errors when operating in a dusty environment \cite{afzalaghaeinaeini_design_2022, phillips_when_2017}. In contrast, heterogeneous failures may encompass distinct types of malfunctions occurring under varying environmental or operational conditions—for example, a sensor failure during one task \cite{schichler_robust_2025}, and motor overheating during another \cite{jorgensen_thermal_2019}. Understanding the patterns of failures is important, as user satisfaction is influenced not only by the robot’s immediate behaviour but also by the history of its interactions with the user \cite{das_explainable_2021, khanna_effects_2023, chang_sorry_2024}.

Robot failures play a significant role in shaping human perception, particularly with respect to trust and perceived intelligence. Trust is a key factor in human–robot collaboration \cite{zhang_can_2025}, as it influences not only the user’s willingness to engage with the robot but also their tolerance for its mistakes \cite{salem_evaluating_2015, hancock_meta-analysis_2011, mirnig_err_2017}. Perceived intelligence, meanwhile, reflects whether the robot is seen as competent and capable of handling the task at hand, serving as one of the foundations on which trust is built \cite{bartneck_measurement_2009,  lee_trust_2004, mayer_integrative_1995}. Although robot failures can undermine both trust and perceived intelligence, prior research \cite{zhang_sorry_2023, sebo_i_2019, lemasurier_reactive_2024} suggests that robots are capable of regaining them to some extent, particularly when appropriate trust repair strategies are employed, such as offering explanations for the failure \cite{bai_effects_2024}. Trust repair approaches vary in effectiveness: minor failures are often easily forgiven \cite{salem_would_2015, pompe_robot_2022}, whereas high-stakes failures, those that disrupt tasks or compromise safety, tend to permit only partial trust recovery \cite{esterwood_repairing_2025}. Awareness plays an enabling role in trust repair: a robot should first be aware of its failures in order to choose and deliver an appropriate recovery strategy. In some instances, robots may rely on internal diagnostic systems to explicitly detect specific errors \cite{cully_robots_2015}. In other cases, they should infer the occurrence of a failure by interpreting social cues from the user, such as gaze patterns \cite{tabatabaei_gazing_2025} or facial expressions \cite{stiber_using_2023}. This awareness can vary considerably, ranging from full recognition of the failure and its cause to a general sense that something has gone wrong or even complete unawareness. The degree of failure awareness influences not only the robot’s choice of repair strategy but also how intelligent and trustworthy it appears to the user.


In light of these considerations, this study examined how user trust and perceived robot intelligence evolve across repeated failure episodes. Our investigation focused on three components. First, we compared two types of failure sequences, homogeneous sequences (repeated failures of the same type) and heterogeneous sequences (failures of varying types). Second, we explored how the robot’s trust-repair strategy interacts with these failure patterns. Third, we assessed how failure severity (low vs. high) shapes user perceptions across repeated failure episodes. Our findings indicate that heterogeneous failure sequences reduce perceived intelligence more than homogeneous sequences, particularly at the second failure. When failures are highly noticeable, users expect the robot to demonstrate awareness by acknowledging the issue or taking corrective action. Finally, user perceptions of the robot are influenced by the severity of both current and previous failures, especially in relation to trust.

\section{RELATED WORK}

\subsection{Types and Characteristics of Robot Failures}

Robot failures have been extensively studied in the literature, resulting in various classification frameworks aimed at identifying their underlying causes. Carlson et al. \cite{carlson_follow-up_2004} initially proposed a division between: \emph{physical failures}, involving faults in the robot’s hardware or software, and \emph{human failures}, stemming from design flaws or errors in human–robot interaction. Expanding on these definitions, Honig and Oron-Gilad \cite{honig_understanding_2018} introduced a more nuanced framework that distinguishes between \emph{technical failures}, including hardware malfunctions and software bugs, and \emph{interaction failures}, which encompass human-related mistakes, violations of social norms, and context-based ambiguities. More recently, Tian et al. \cite{tian_taxonomy_2021} suggested a taxonomy comprising \emph{performance errors}, which impair task execution (e.g., misinterpreting commands, navigation faults), and \emph{social errors}, which undermine socio-affective competence by breaching interactional expectations (e.g., interrupting inappropriately). In this study, we build on these frameworks by focusing specifically on performance-oriented technical failures.

Building on these taxonomies, it is important to consider how failure types manifest in specific robotic domains, such as manipulation. Failures in robotic manipulation present unique challenges, as tasks ranging from simple grasping to complex pick-and-place operations often expose system vulnerabilities. In such systems, typical failure types include hardware faults (e.g., overheating or servo motor overloading) \cite{zhang_fault_2025}; control and actuation deficiencies \cite{sabry_review_2024}; perception errors, such as target mislocalisation or occlusion \cite{zhu_failure_2021}; grasping challenges, including object slippage and unstable grips \cite{hu_learning_2024}; and plan-execution breakdowns that hinder successful task completion \cite{k_learning_2024, hietanen_benchmarking_2021}.

In parallel, research has examined how specific characteristics of robot failures influence users’ trust, perceived intelligence, and behavioural responses during interaction. Failure characteristics such as type \cite{zhang_sorry_2023, tabatabaei_gazing_2025}, severity \cite{van_waveren_take_2019, gideoni_is_2024}, frequency \cite{esterwood_three_2023, nesset_robot_2023}, and timing \cite{desai_impact_2013, morales_interaction_2019} have all been shown to influence user attitudes and behaviour. For instance, Stiber et al. \cite{stiber_not_2021} observed that more severe failures elicit faster user reactions and heightened socially expressive responses, such as verbal exclamations, smiling, or making exaggerated facial expressions. Similarly, Garcia et al. \cite{garcia_goo_being_2025} demonstrated in navigation contexts that the type of failure differentially influences trust, with personal-space violations leading to steeper declines than abrupt stops or corner-turning errors. Beyond HRI, psychological research indicates that preceding events shape the impact of subsequent ones: a severe negative event following a milder one can heighten expectancy violation and amplify its negative effect on perceptions \cite{baumeister_bad_2001}. Taken together, these findings suggest not only that severity intensifies user responses but also that sequencing may matter, underscoring the need to examine their joint influence in HRI.

\subsection{Trust and Failure in Human–Robot Interaction}

Trust is a foundational element in effective HRI, influencing whether users choose to rely on, follow, or collaborate with robotic systems \cite{hancock_meta-analysis_2011, mayer_integrative_1995, lee_trust_2004}. Trust in HRI is typically defined as the user’s belief in a robot’s ability to perform competently and reliably in its operational context, especially when outcomes are uncertain or risky \cite{sitkin_explaining_1993, lee_trust_2004}. Trust is dynamic and context-sensitive \cite{kelch_would_2024, duan_trusting_2025}; it builds through consistent, reliable behaviour and can degrade rapidly when the robot fails to meet user expectations \cite{desai_impact_2013, dzindolet_role_2003, schaefer_meta-analysis_2016}.

Over time, users’ trust in robotic systems can develop through sustained interaction and experiential learning, even in the absence of perfect performance. For instance, Alzahrani et al.~\cite{alzahrani_what_2025} found that even with a constant error rate, participants reported increased trust over multiple sessions with a NAO robot, as indicated by both self-reports and behavioural cues such as smiling and reduced vocal pitch variability. Similarly, Benford et al.~\cite{benford_somatic_2025} observed that professional dancers, initially confronted with a robot exhibiting unpredictable behaviour, gradually developed trust as they became skilled at anticipating its unique patterns. In a comparable study, Kahr et al.~\cite{kahr_good_2025} found that logistics professionals’ trust in an AI planning system improved over time. Importantly, this occurred despite the system’s imperfect performance, as users learned to anticipate its suggestions and make minor adjustments as needed. Yet, while trust can grow through repeated interaction, it can also be lost quickly when failures are less predictable. Trust can disappear fast when something goes wrong, especially if the robot behaves unexpectedly. Nesset et al.~\cite{nesset_robot_2023} demonstrated that trust tends to decline most sharply following the initial failure, with subsequent failures less impactful. These findings highlight predictability as a critical factor in trust development, suggesting that repeated exposure to the same type of failure may be easier for users to adapt to, whereas varied and less predictable failure patterns pose greater challenges for trust calibration.

\subsection{Trust Repair Strategies}

Previous research across both HRI and human–AI domains has demonstrated that trust can be partially restored through specific behaviours following a failure. A variety of trust repair strategies have been explored in this context, including offering apologies~\cite{mahmood_owning_2022, pareek_its_2025, esterwood_theory_2023, sebo_i_2019}, providing explanations~\cite{lyons_explanations_2023, kox_trust_2021}, making promises to improve future performance~\cite{pareek_its_2025, esterwood_theory_2023}, and even denying responsibility~\cite{sebo_i_2019, zhang_sorry_2023}. Among these, explanations, apologies, and promises of improvement have consistently been found to be the most effective strategies for trust restoration.



LeMasurier et al.~\cite{lemasurier_reactive_2024} showed that explanations after a failure can restore user trust and improve perceptions of the robot’s intelligence, with detailed and specific explanations proving more effective than generic acknowledgements. Extending this line of inquiry, Kraus et al.~\cite{kraus_sorry_2023} found that explanations were even more effective when combined with an apology. However, Wachowiak et al.~\cite{wachowiak_when_2024} showed that user expectations of appropriate robot responses not only differ across contexts such as errors, uncertainty, and inability, but also vary within each context.

Promises have been also identified as a potentially effective strategy for trust repair; however, their efficacy is highly context-dependent and may backfire in the presence of repeated failures. Karli et al.~\cite{karli_what_2023} investigated the impact of a robot’s promise to enhance future performance in comparison to a simple failure explanation. Their findings indicated that promises resulted in a significantly greater increase in users’ trust. Conversely, Nesset et al.~\cite{nesset_robot_2023} demonstrated that when a robot issued a promise but subsequently failed again, the resulting decline in trust was more pronounced than when the robot offered no response to its failure.

While much of the existing research on trust repair in human-robot interaction has emphasised verbal strategies, recent studies have begun to explore more embodied, action-oriented approaches. Lane et al.~\cite{lane_robots_2024}, for instance, demonstrated that robots which autonomously corrected their own mistakes, without seeking user permission, were more effective in restoring trust than those that either waited for user approval before acting or showed no reaction to the failure. Building on this, in an online study, Lee et al.~\cite{lee_rex_2024} demonstrate that users report significantly greater trust when a robot attempts automated repair, even if imperfect, compared to when it remains inactive.


Previous research has demonstrated that combining an explanation of a failure with an apology is among the most effective strategies for repairing trust in human–robot interaction. In addition, physical repair actions have been identified as a promising approach for restoring trust. However, a critical gap remains. Because robot failures can stem from diverse causes, they may be fully detectable, partially detectable, or not detectable at all, meaning that a robot’s level of failure awareness can vary substantially. Moreover, existing work suggests that the effectiveness of both verbal and physical repair may be context-dependent, varying with the type of robot failure. These considerations motivated our investigation into how participants’ trust is influenced by different levels of failure awareness, particularly in a human–robot collaboration setting.

\section{Methodology}

\subsection{Research Questions}


Although some studies have examined how users respond to repeated robot failures, it remains unclear whether users react differently to homogeneous versus heterogeneous sequences of failures. Research has also explored trust-repair strategies, yet little is known about how their effectiveness depends on both the type of failure and the robot’s level of awareness. Finally, while prior work has highlighted the role of failure severity, it has not fully considered how the severity of one failure interacts with that of preceding failures to shape user perceptions. Building on these gaps, we propose the following research questions:

\begin{itemize}


  \item \textbf{RQ1:} \label{RQ1} Does the pattern of repeated failures, homogeneous (same type) vs. heterogeneous (mixed types), affect how trust and perceived intelligence evolve over time?


  \vspace{0.5\baselineskip}

  \item \textbf{RQ2:} \label{RQ2} How do different trust-repair strategies, based on the robot’s level of failure awareness, interact with failure type (freezing failure, planning failure, grasping failure) to influence user trust and perceived intelligence?

  \vspace{0.5\baselineskip}

  \item \textbf{RQ3:} \label{RQ3} How do high-severity and low-severity robot failures influence users’ trust and perceived intelligence, as influenced by the severity of the preceding failure?

\end{itemize}

\subsection{Experimental Design}

\begin{figure*}[thb]
  \centering
  \begin{subfigure}[t]{0.31\linewidth}
    \centering
    \includegraphics[width=\linewidth]{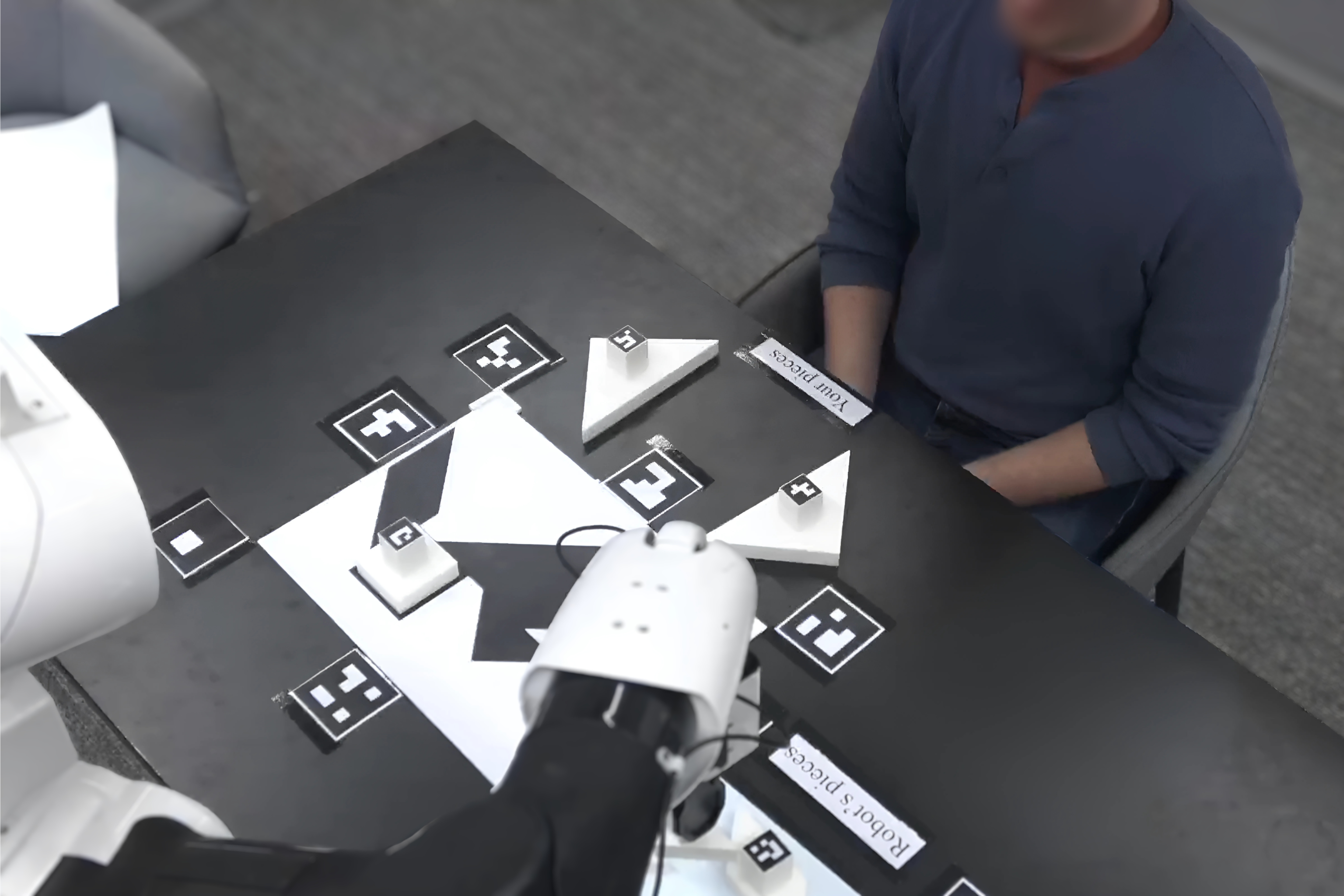}
    \caption{Freezing failure: the robot pauses mid-action while holding a puzzle piece.}
    \Description{A person sits at a table across from a robotic arm during a Tangram puzzle task. The tabletop shows the puzzle silhouette and its pieces with ArUco markers, with labelled areas reading “Your Pieces” and “Robot’s Pieces.” The robot’s gripper is positioned near its pieces area, where it has frozen.}

    \label{fig:failure-freezing}
  \end{subfigure}
  \hfill
  \begin{subfigure}[t]{0.31\linewidth}
    \centering
    \includegraphics[width=\linewidth]{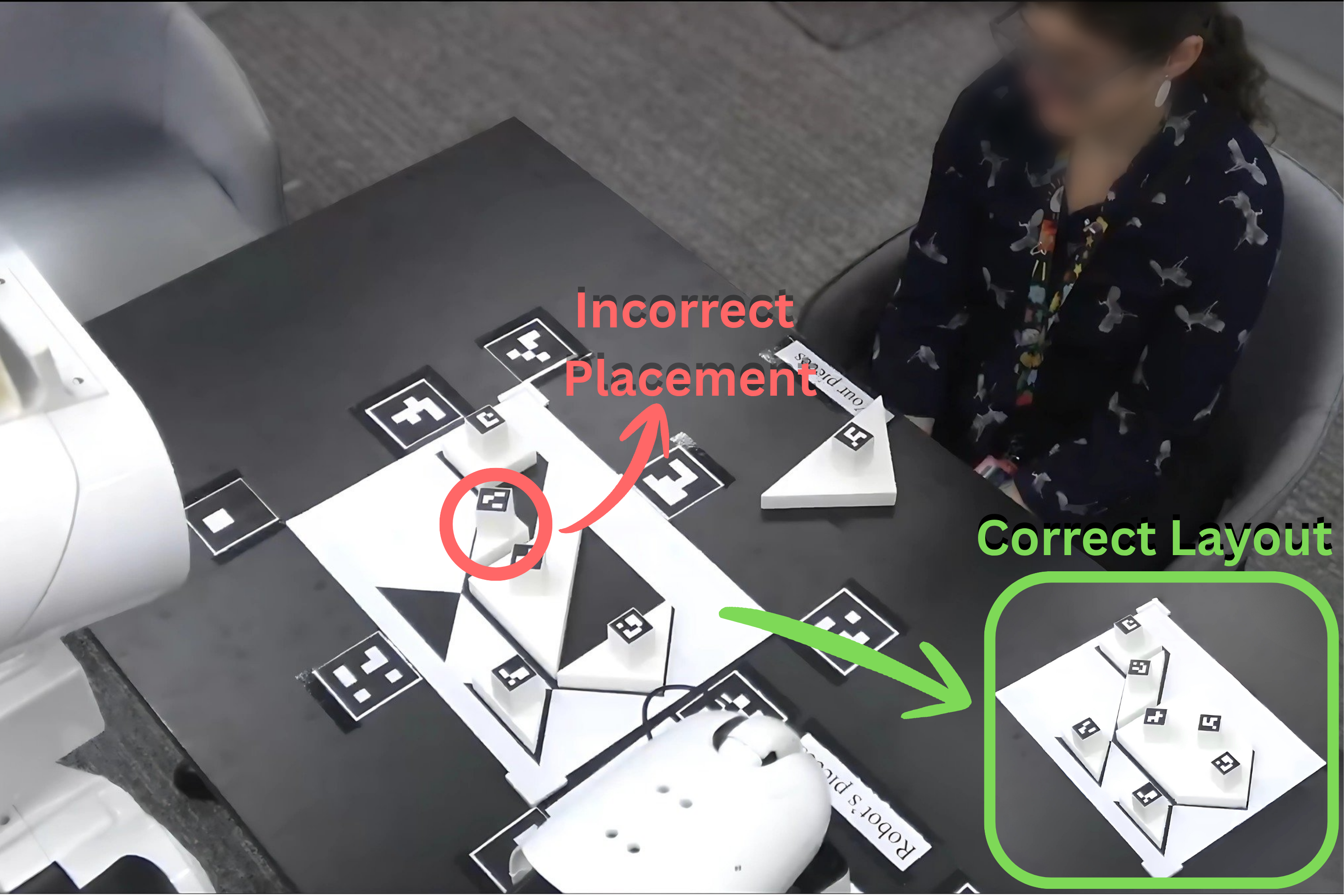}
    \caption{Planning failure: the robot places a triangle in the slot intended for a parallelogram.}
    \Description{A person sits at a table across from a robotic arm during a Tangram puzzle task. The puzzle pieces, marked with ArUco tags, are being placed on the silhouette board. A red annotation highlights an incorrect placement of one piece, while a green annotation shows the correct layout in a reference image.}
    \label{fig:failure-planning}
  \end{subfigure}
  \hfill
  \begin{subfigure}[t]{0.31\linewidth}
    \centering
    \includegraphics[width=\linewidth]{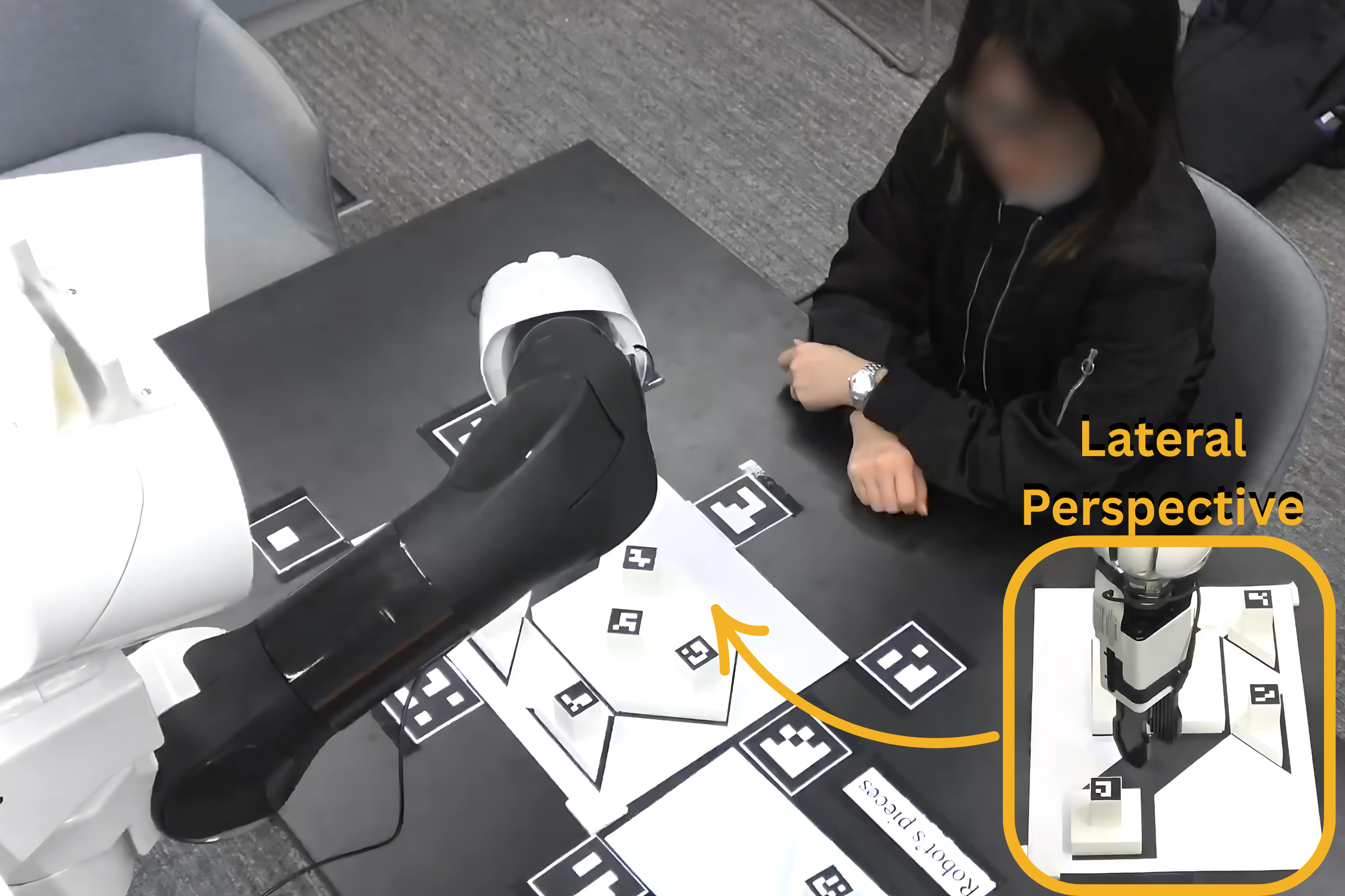}
    \caption{Grasping failure: the robot fails to pick up the piece due to an incomplete gripper closure}
    \Description{A person sits across from a robotic arm during a Tangram puzzle task. The robot’s gripper is empty, with no piece being held. A highlighted inset labelled “Lateral Perspective” shows a side view of the gripper making contact with the puzzle silhouette paper.}
    \label{fig:failure-grasping}
  \end{subfigure}
  \caption{Three failure types used in the study, shown as screenshots from the puzzle assembly task.}
  \label{fig:failure-types}
\end{figure*}

We employed a collaborative Tangram puzzle task previously introduced in human-robot interaction research \cite{tabatabaei_gazing_2025, kirschner_yumi_2016, correia_exploring_2018}. In this task, a human participant and a robot worked together to solve six Tangram puzzles. The robot was responsible to place four pieces (two small triangles, a square, and a parallelogram), while the participant handled the remaining three pieces (two large triangles and one medium triangle). During the collaboration, the robot intentionally failed. The type of failure varied, and the robot’s reaction to its failure was systematically manipulated based on its level of awareness.

\subsubsection{Robot Failures}
This study focused on common failure types in such settings, all of which are characterised by the absence of any potential harm to the object or the user.
We implemented three types of failures for the robot: freezing, planning, and grasping (Figure \ref{fig:failure-types}). 

\begin{itemize}
\item \textbf{Freezing Failure}. Similar to failures reported in prior work \cite{fan_getting_2019, ramesh_robot_2022}, this refers to a temporary pause in the robot’s operation, often due to limited processing capacity or system overload. In our experiment, the robot was programmed to freeze for 15 seconds, consistent with the duration used in prior studies \cite{maehigashi_effect_2024, kontogiorgos_embodiment_2020}, while holding an object in its end-effector—specifically after picking up a piece and before placing it. This failure was systematically introduced during the placement of the robot’s second puzzle piece.

\item \textbf{Planning Failure}. Following approaches in earlier studies \cite{k_learning_2024}, this failure involved the robot placing an object incorrectly, typically resulting from inaccurate perception or target misjudgment. In our task, the robot was programmed to place a small triangle in the position designated for the parallelogram. This failure was systematically introduced during the placement of the robot’s third puzzle piece.

\item \textbf{Grasping Failure}. As observed in prior research \cite{inceoglu_fino-net_2021}, this failure refers to unsuccessful object grasping, often caused by gripper malfunctions or inaccurate force sensing. In our experiment, the robot was programmed not to fully close its gripper, resulting in a failed grasp. Despite this, the robot continued moving as if it were holding the puzzle piece. This failure was systematically introduced during the placement of the robot’s fourth puzzle piece.
\end{itemize}

\subsubsection{Robot Awareness of Failures}
In addition to manipulating the type of failure, we varied the robot’s reaction to its own failure. Participants experienced three levels of robot awareness. A detailed overview of the robot’s behaviour, depending on its failure type and level of awareness, is presented in Table~\ref{TableExplanations}.

\begin{itemize}
\item \textbf{No Awareness.} The robot fails to recognise the error and continues its actions as if no mistake has occurred, providing no explanation or indication of the failure. This leaves the participant responsible for identifying and addressing the issue, if necessary.

\item \textbf{Partial Awareness.} After returning its arm to the initial position following a mistake, the robot acknowledges that something might have gone wrong but expresses uncertainty about the specific issue. It prompts the participant to intervene if necessary.

\item \textbf{Full Awareness.} After returning its arm to the initial position following a mistake, the robot accurately identifies the nature of the failure and provides a verbal explanation to the participant, describing what has gone wrong. For the planning and grasping failures, the robot also initiates corrective actions to resolve the issue and continue with the task.
\end{itemize}

\begin{table}[ht]
\centering
\caption{Robot explanations based on failure type and awareness level}
\Description{Robot explanations vary by awareness level and failure type. With no awareness, the robot gives no explanation or fix attempt. With partial awareness, it issues a generic apology without repair. With full awareness, explanations are specific to the failure type and include a fix attempt.}
\label{TableExplanations}

\renewcommand{\arraystretch}{1.5}
\begin{tabular}{p{1.5cm} p{1.3cm} p{0.6cm} p{3.5cm}}
\hline
\textbf{Awareness Level} & \textbf{Failure Type} & \textbf{Fix} & \textbf{Robot Explanation} \\
\hline

No \text{Awareness}
& Freezing/ Planning/ Grasping 
& No & No explanation or acknowledgement. 
 \\
\hline

Partial Awareness 
& Freezing/ Planning/ Grasping 
& No & \textit{``Something might have gone wrong with me. I apologise if it caused any trouble. If any fixes are needed, please make them.''}
 \\
\hline
& Freezing
& Yes & \textit{``Due to some technical issues, I froze for a few seconds. I apologise for that.''}
 \\
Full \text{Awareness}
& Planning
& Yes & \textit{``I made a wrong decision and placed the object in the wrong location. I apologise for that.''}
 \\

& Grasping 
& Yes & \textit{``I didn’t fully close my gripper, so I couldn’t place the object. I apologise for that.''}
 \\
\hline

\end{tabular}
\end{table}

\subsubsection{Robot Failure Sequence Patterns}

To examine how the sequence of failures shapes users’ perceptions of a robotic teammate, the study compared two conditions: a homogeneous sequence and a heterogeneous sequence. In both conditions, participants experienced three failures during a collaborative task, occurring at predetermined points in the interaction (Puzzles 2, 4, and 6). 

\paragraph{Homogeneity Condition}
In this condition, each participant encountered the same type of failure across all three failure episodes. For example, one participant might experience the freezing failure in Puzzles 2, 4, and 6, while another might encounter only the grasping or only the planning failure at those same points. Although the specific failure type differed across participants, it remained consistent within each participant’s experience. This design simulated scenarios in which, through repeated exposure, the failure became familiar and attributable to a single underlying cause.

\paragraph{Heterogeneity Condition}
In this condition, the participant experienced three different failure types, with each type occurring only once—in Puzzles 2, 4, and 6. For example, a participant might encounter a freezing failure in Puzzle 2, a grasping failure in Puzzle 4, and a planning failure in Puzzle 6. This design introduced variability and reduced predictability in the robot’s behaviour, simulating a diffuse, system-wide reliability issue in which failures could not be attributed to a single recurring cause.

\subsubsection{Design Structure}
The experiment was designed around two independent variables: the failure sequence type (homogeneous vs. heterogeneous) and the robot’s level of failure awareness (no awareness, partial awareness, full awareness). We employed a mixed-design approach, in which the failure sequence type served as a between-subjects factor, while the robot’s level of failure awareness was manipulated as a within-subjects factor. Participants’ perceived failure severity was assessed based on their responses collected during the post-experiment phase.

The first puzzle contained no failures, allowing participants to become familiar with the task, observe the robot’s normal behaviour, and develop an accurate mental model of its capabilities \cite{stiber_effective_2022}. To ensure that failures were evenly distributed throughout the session, the task was structured so that participants alternated between puzzles with and without a failure. To minimise potential order effects, both the sequence of robot awareness conditions and the order of failure types were counterbalanced across participants.

\subsection{Participants}


A total of 54 participants were recruited to ensure full coverage of all combinations of failure sequences and trust repair strategies used in the study. Participants were recruited through the university’s participant recruitment platform and included 39 female, 14 male, and 1 non-binary individual. Eighteen participants were assigned to the homogeneous condition, with six participants allocated to each failure type. The remaining 36 participants were assigned to the heterogeneous condition, in which all three failure types were presented.

Participants ranged in age from 18 to 39 years ($M = 25.59$, $SD = 5.45$) and were primarily students or university staff. None reported prior experience working with robots or formal training in programming. All participants possessed advanced or proficient English-language proficiency. Each participant received a gift voucher as compensation for their time.

The study was conducted in a laboratory facility at the University of Melbourne. The study was approved by the University of Melbourne Human Research Ethics Committee (Approval No. 2025-29401-62628-4). At the end of the session, participants were fully debriefed and informed that the study aimed to investigate user responses to robot failures during collaborative interaction.

\subsection{Apparatus and Materials}

To make the task suitable for collaborative interaction between the robot and the participant, we developed a large-scale physical version of the Tangram puzzle. The seven Tangram pieces were 3D-printed using white PLA material. To facilitate robotic manipulation, a small cube was attached to the top of each piece, enabling the robot to grip them more reliably. Additionally, an ArUco marker was affixed to the top of each cube, allowing the robot to accurately detect and localise the pieces. Each target shape for the puzzle was printed on A3-sized paper, providing a physical template. Participants and the robot were required to place the Tangram pieces directly onto the printed outlines to complete the puzzle collaboratively. The robot’s pieces (two small triangles, one square, and one parallelogram) were placed on the same table near the robot, while the participant’s pieces (two large triangles and one medium triangle) were positioned directly in front of them.

The robot used in this study was the TIAGo robot developed by PAL Robotics. It features a single arm with a parallel gripper as its end-effector. To enhance the robot’s ability to pick up objects with better precision, a camera was mounted on the robot’s end effector. The robot was programmed using the Robot Operating System (ROS) version 1. Additionally, to generate more natural and human-like vocal output, all spoken messages were preprocessed into audio files using the text-to-speech service provided by \href{https://luvvoice.com/}{Luvvoice}.

To enable the robot to accurately detect and localise Tangram pieces during the task, we developed a custom ROS node written in Python. This node uses the "tf" library to manage coordinate transformations between the robot's camera frame and its base frame. Upon receiving the position and orientation data of the Tangram pieces, obtained via ArUco markers affixed to each piece, the system utilised the tf library to transform the detected poses from the camera frame into the robot’s base frame, thereby enabling accurate localisation for subsequent manipulation. The robot was programmed to operate almost fully autonomously throughout the task, requiring only minimal input to specify when and which object shape to place.

To facilitate precise and coordinated manipulation during the Tangram task, the robot's arm was controlled through a combination of inverse kinematics (IK), trajectory generation, and ROS-based action interfaces. The robot’s URDF model was used to construct a kinematic chain from the base (\texttt{torso\_lift\_link}) to the end-effector using the KDL library. To position the robot’s gripper at a specific location and orientation in space, the required joint angles of the arm were computed using a Levenberg--Marquardt-based inverse kinematics solver (\texttt{ChainIkSolverPos\_LMA}). To enable smooth and natural movements, cosine interpolation was applied between the current and target joint configurations, generating a trajectory composed of 100 intermediate waypoints. These trajectories were published directly to the arm controller.

\subsection{Procedure}

The experiment was conducted in a laboratory environment at a university. Upon arrival, participants were greeted by the experimenter and briefed on the purpose of the study and the nature of the interaction. They were informed that the task involved collaboratively solving six Tangram puzzles with a robot, taking turns to place puzzle pieces. Participants signed a consent form and were told that the robot was autonomous. They were advised that the robot might occasionally make minor errors during the task. Additionally, they were informed that the robot had some capacity to detect its own mistakes, and that when a failure was fully detected, it would attempt to correct it. The sequence of puzzle shapes presented included: Rocket, Turtle, Cat, Rabbit, Dog, and Swan. Participants were instructed to wait for the robot’s verbal cue, ``Now it is your turn'', before placing any pieces and to avoid touching any objects while the robot was performing its actions. Participants were not time-limited during their turns and were free to place their pieces wherever they chose.

\begin{figure*}[htbp]
  \centering
  \includegraphics[width=\linewidth]{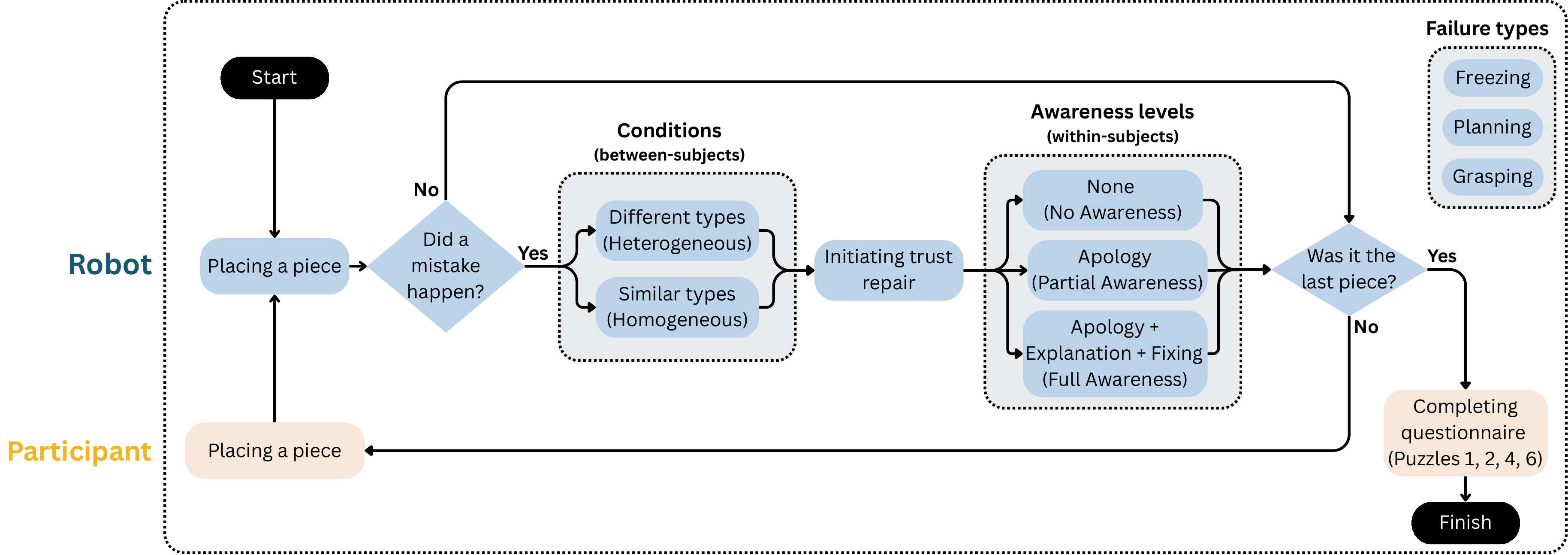}
  \caption{Flowchart depicting the collaborative process of solving a single puzzle between a robot and a human participant.}
  \Description{Flowchart showing the study sequence: start -> placing puzzle pieces (by robot) -> error check (by robot). If no error, the task continues. If an error occurs, it is categorised as freezing, planning, or grasping. Robot awareness shapes the response: no awareness (no explanation), partial awareness (apology only), or full awareness (apology, explanation, and attempted fix). The process repeats until the final piece is placed, after which the participant completes a questionnaire.}
  \label{fig:FlowChart}
\end{figure*}

They were not informed about the number or nature of the robot’s potential errors. Prior to each puzzle, participants were given one minute to observe the silhouette and plan their moves. The sequence of events in a typical puzzle round is illustrated in Figure \ref{fig:FlowChart}.

The experimenter left the room during each puzzle and returned only after its completion to change the puzzle silhouette paper for the next round. The robot was remotely monitored and controlled by the experimenter from an adjacent control room.

Each Tangram puzzle featured a distinct silhouette, resulting in different target placements for the robot’s pieces. These placements were predefined in the robot’s program prior to the experiment. At the start of each trial, the robot localised its pieces using its head-mounted camera. To enhance accuracy during object manipulation, it used its end-effector-mounted camera to fine-tune its gripper position before picking up each piece. The robot consistently started from an initial posture, moved above the selected puzzle piece, grasped it using the gripper, and transported it to its designated location within the silhouette.

During the one-minute planning period, the robot identified the positions of its pieces and prepared for manipulation. Once the one minute had elapsed, it announced, ``Let’s start solving the puzzle,'' and placed its first piece. It then signalled the participant to proceed. The robot followed a fixed order when placing its pieces: beginning with the square, followed by the small triangles, and concluding with the parallelogram.
In instances where a participant mistakenly placed a piece in the wrong location, the robot did not attempt to correct the error, and remained silent. If the participant placed the object in a location that belonged to the robot’s next piece, the robot skipped that piece in the sequence and instead placed the following one; otherwise, it followed the predefined piece placement order. This design ensured that participants’ evaluations remained consistent and were not biased by the robot correcting their mistakes.

Participants completed questionnaires assessing their trust and perceived intelligence of the robot after Puzzle 1 (an error-free baseline) and after Puzzles 2, 4, and 6, each of which incorporated pre-programmed robot failures. Following the final puzzle, participants also responded to a series of open-ended questions (listed in  Table~\ref{TabelQual}) to further elaborate on their experiences.

The full session, which involved collaboratively completing six puzzles, lasted approximately 30 minutes. Following the sixth puzzle, participants spent up to an additional 30 minutes responding to the open-ended questions.

\subsection{Measures}

\subsubsection{Quantitative Measures and Analysis}

To measure user trust, we employed four items from the performance subscale of the Multi-Dimensional Measure of Trust (MDMT v2) \cite{ullman_mdmt_2023}. These items asked participants to rate the robot on how reliable, consistent, competent, and skilled it appeared. Ratings were provided on a 7-point Likert scale. For perceived intelligence, we selected three items, responsible, intelligent, and sensible, from the Godspeed Questionnaire Series \cite{bartneck_measurement_2009}, rated on a 5-point Likert scale. Internal consistency of the scales was assessed using Cronbach’s alpha. The trust scale demonstrated excellent reliability ($\alpha = 0.89$), and the perceived intelligence scale showed acceptable reliability ($\alpha = 0.77$).

We designed three cumulative link mixed models (CLMMs), each addressing one of the research questions. All models were estimated using the \texttt{clmm()} function from the \texttt{ordinal} package in R. The dependent variable was the ordinal rating of either trust or perceived intelligence reported by each participant after a robot failure. Random intercepts were specified for participants and for scale items (i.e., individual questionnaire items) to account for clustering. Predictor significance was evaluated using Type~II likelihood–ratio $\chi^2$ tests (via \texttt{RVAideMemoire::Anova}), and estimated marginal means (EMMs) with Tukey-adjusted pairwise contrasts were obtained using the \texttt{emmeans} package on the latent scale.

\paragraph{Model 1: Impact of Failure Sequence Structure (Homogeneous vs. Heterogeneous).}\label{model2method}
To examine how homogeneous versus heterogeneous sequences influenced participants’ perceptions of trust and perceived intelligence, we estimated a CLMM. Model~1 specified fixed effects for the ordinal position of the failure, the sequence condition (homogeneous vs.\ heterogeneous), and their interaction. The robot’s current and previous failure awareness were included as control variables to account for their potential influence, and participants’ pre-failure rating was added as a covariate.

\paragraph{Model 2: Effects of Failure Awareness Across Failure Types.}\label{model3method}

To investigate the effectiveness of recovery strategies across different types of robot failures, we estimated a second CLMM. Model~2 included fixed effects for the type of current failure, the robot’s current failure awareness, and their interaction. Preceding failure type and preceding failure awareness were included as control variables, along with two additional covariates: participants’ pre-failure rating and the ordinal position of the failure.

\paragraph{Model 3: Sequential Effects of Failure Severity.}\label{model1method}

To examine how the severity of the current failure, together with the severity of the preceding failure, influenced participants’ perceptions, we estimated a third CLMM. Failure severity was operationalised using participants’ rankings of the three distinct failure types in the heterogeneity condition (Question 1, Table \ref{TabelQual}). These rankings were then aggregated to produce a stimulus-level severity score for each failure type.
Model~3 specified fixed effects for the severity of the current failure, the severity of the preceding failure, and their interaction. Participants’ pre-failure rating and the ordinal position of the failure (second vs.\ third) were included as covariates, and the robot’s verbal reactions were included as control variables to account for their potential influence.

To visualise the model predictions, we plotted the expected trust and perceived intelligence scores for each condition. These were computed as probability-weighted averages of the ordinal response scale values (1–7 for trust and 1–5 for perceived intelligence) derived from the CLMM predictions. Error bars represent 95\% confidence intervals.

\subsubsection{Qualitative Measures and Analysis}\label{Qual measures}


At the conclusion of the sixth puzzle, participants responded to a series of open-ended questions (see Table~\ref{TabelQual}). The first question explored the reasons participants perceived certain failures as more severe than others, highlighting the features that informed their judgments. The second question examined how the robot’s reactions to its own mistakes shaped participants’ perceptions of the interaction. Responses were analysed using reflexive thematic analysis \cite{terry_summary_2021} to identify themes in participants’ severity judgments and their interpretations of the robot’s mistake awareness and responses during collaboration.

\begin{table}[htb]
\centering
\caption{Post-task interview questions based on participants’ condition (homogeneity vs. heterogeneity).}
\Description{Post-task interview questions for participants in the homogeneity and heterogeneity conditions. Participants in the homogeneity condition were asked about the perceived severity of a repeated mistake. Participants in the heterogeneity condition were asked to compare the severity of three different mistakes. All participants were asked how the robot’s reaction to its mistakes influenced their experience.}
\label{TabelQual}

\renewcommand{\arraystretch}{1.5}
\begin{tabular}{p{0.05\columnwidth} p{0.23\columnwidth} p{0.57\columnwidth}}
\hline
\textbf{No.} & \textbf{Condition} & \textbf{Question} \\
\hline

1 & Homogeneity &
\textit{You have observed the robot making the same mistake in three different puzzles. Did you perceive the mistake as severe in this collaboration? Please explain.} \\

1 & Heterogeneity &
\textit{You observed the robot making three types of mistakes during different puzzles: freezing, placing an object incorrectly, and failing to pick up an object. Which mistake did you find the most severe, and which was the least severe? Please explain.} \\

\hline
2 & Homogeneity/ Heterogeneity &
\textit{When the robot made a mistake, how did its reaction (or lack of reaction) affect you? In what ways? (Consider its awareness of the mistake and whether it attempted to fix it afterward.)} \\
\hline

\end{tabular}
\end{table}


In the reflexive thematic analysis, one researcher conducted the initial open coding on approximately 20\% of the responses to generate a preliminary set of descriptive codes. These initial codes and emerging categories were then reviewed and discussed with the co-authors to refine the coding scheme. Using this refined scheme, the primary researcher coded the remaining responses and iteratively organised the codes into higher-level themes. The final themes were subsequently reviewed and confirmed by the co-authors.

\section{RESULTS}


\subsection{Quantitative Results}

The following sections present the quantitative analyses of participants’ perceptions of the robot, focusing on trust and perceived intelligence. Using CLMMs and Type~II likelihood–ratio $\chi^2$ tests, we examined how these perceptions were shaped by different characteristics of robot failures, including their sequence, the robot’s awareness of its own errors, and severity.

\subsubsection{Impact of Failure Sequence Patterns}

This section reports the results of the Type~II likelihood–ratio $\chi^2$ tests conducted on Model-1, as described in section~\ref{model2method}. The analysis examines how participants’ perceptions of the robot, specifically trust and perceived intelligence, change over the course of the interaction as they encounter multiple failures, with a focus on the effects of different failure sequence patterns. This section addresses \hyperref[RQ1]{RQ1}.

\begin{figure}[tb]
  \centering
  \begin{subfigure}[tb]{0.48\linewidth}
    \centering
    \includegraphics[width=\linewidth]{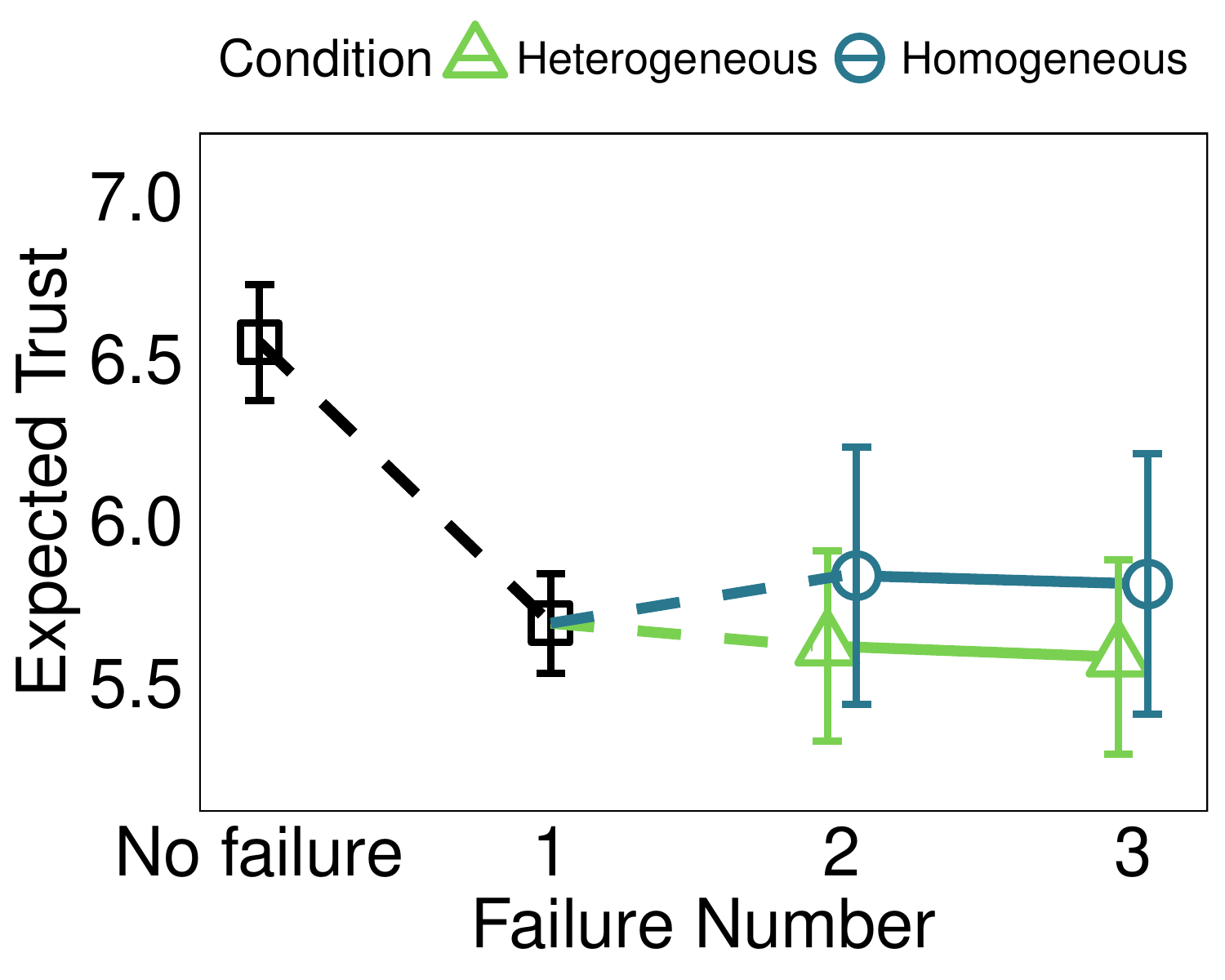}
    \caption{Trust ratings}
    \Description{The figure is a line graph showing expected trust ratings on the y-axis (ranging from 5.5 to 7.0) and failure number on the x-axis (“No failure,” 1, 2, and 3). Two conditions are plotted: homogeneous failures and heterogeneous failures. Both lines start near 6.5 at “No failure” and decline to around 5.7 after the first failure. At the second failure, the heterogeneous condition drops further to about 5.6, while the homogeneous condition remains around 5.8. At the third failure, both conditions remain approximately at their second-failure levels.}

  \end{subfigure}
  \hfill
  \begin{subfigure}[tb]{0.48\linewidth}
    \centering
    \includegraphics[width=\linewidth]{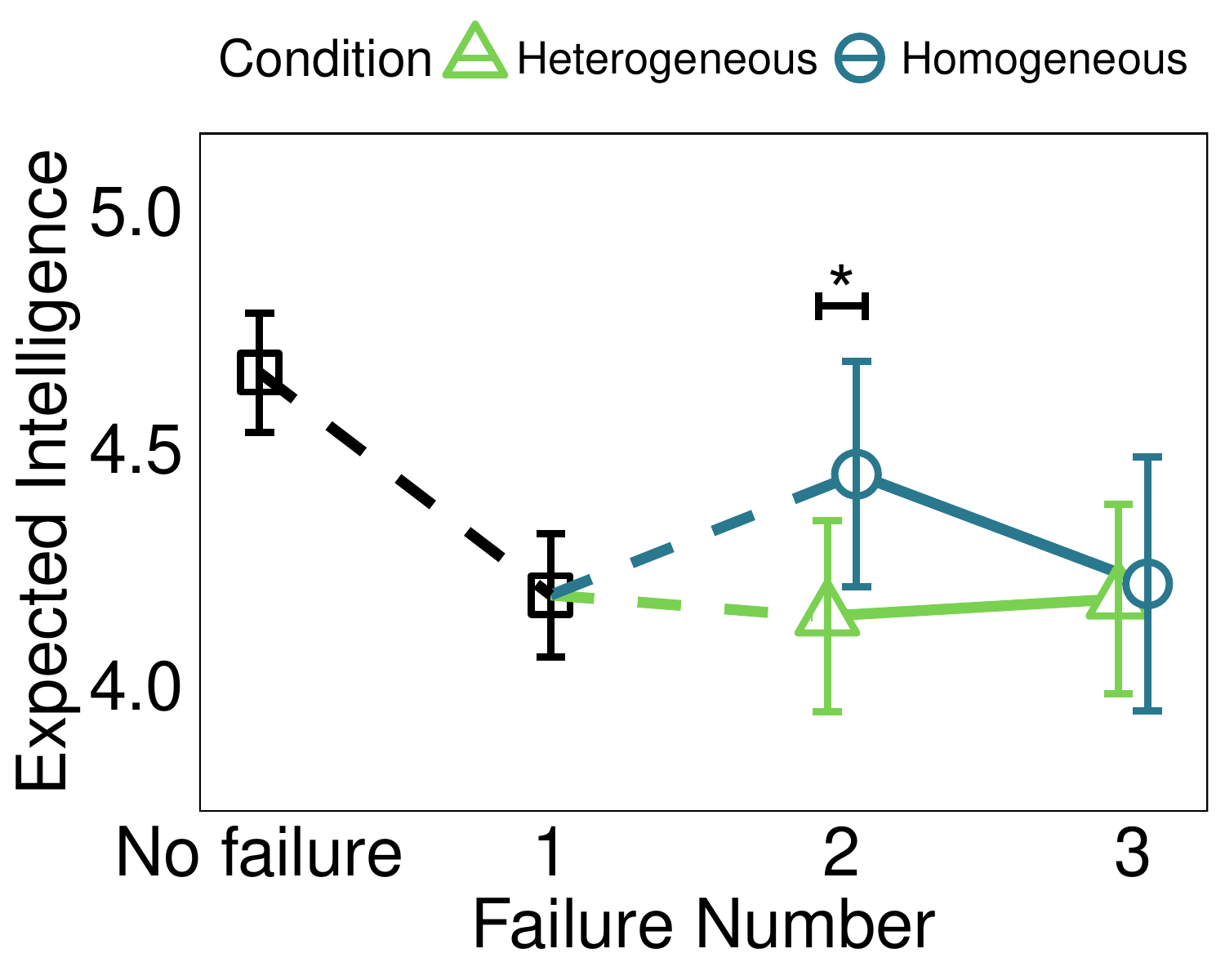}
    \caption{Perceived intelligence ratings}
    \Description{The figure is a line graph showing expected perceived intelligence ratings on the y-axis (ranging from 4.0 to 4.5) and failure number on the x-axis (“No failure,” 1, 2, and 3). Two conditions are plotted: homogeneous failures and heterogeneous failures. Both lines begin near 4.6 at “No failure” and decline to around 4.2 after the first failure. At the second failure, the heterogeneous condition shows no additional drop, while the homogeneous condition remains around 4.5. By the third failure, both lines converge at a value of 4.2. A p-value of less than 0.05 indicates a statistically significant difference between the two conditions at failure number 2.}
  \end{subfigure}
  
  \caption{model-predicted (expected) values of participants’ trust (left) and perceived intelligence (right) across robot failure number (from no failure to the third failure), separated by condition (homogeneity vs.\ heterogeneity). Error bars represent 95\% confidence intervals. Black square markers at the no failure and first failure points represent mean ratings obtained prior to the assignment of participants to the homogeneous and heterogeneous conditions. An asterisk denotes statistical significance at p < .05.}
  \label{fig:trust-intel-failurenumber}

\end{figure}

Following the successful completion of the first puzzle, prior to any robot failures, participants rated their trust in the robot at an average of 6.66 ($SD = 0.65$), and its perceived intelligence at 4.67 ($SD = 0.46$). After the first failure, both ratings declined, with trust dropping to 5.69 ($SD = 0.15$) and perceived intelligence to 4.19 ($SD = 0.13$). Thereafter, the rating patterns began to diverge across failure-sequence conditions. In the homogeneous condition, ratings increased following the second failure, whereas in the heterogeneous condition, they showed a slight further decline. By the third failure, trust decreased slightly in both conditions. For perceived intelligence, however, ratings decreased in the homogeneous condition but increased slightly in the heterogeneous condition, with both conditions ultimately returning to levels comparable to those observed after the first failure. These patterns are shown in Figure~\ref{fig:trust-intel-failurenumber}.

\paragraph{Model-1 Results for Trust.}
The CLMM analysis indicated that \textit{condition} did not have a significant effect, $\chi^2(1) = 0.85$, $p = .36$. Similarly, \textit{failure number} was not a significant predictor, $\chi^2(1) = 0.13$, $p = .72$. The interaction between \textit{failure number} and \textit{condition} was also non-significant, $\chi^2(1) < 0.01$, $p = .97$. Among the covariates, \textit{pre-failure trust} emerged as a significant predictor, $\chi^2(1) = 4.71$, $p < .05$. With respect to the control variables, both \textit{current failure awareness}, $\chi^2(2) = 32.68$, $p < .001$, and \textit{previous verbal reaction}, $\chi^2(2) = 7.61$, $p < .05$, were significant.

Post hoc contrasts did not reveal significant differences between conditions for either failure point. After the second failure, ratings in the heterogeneous condition did not differ significantly from those in the homogeneous condition ($z = -0.85$, $p = .39$). Similarly, after the third failure, no significant difference was observed between conditions ($z = -0.87$, $p = .38$).

\paragraph{Model 1 Results for Intelligence.}
The CLMM analysis showed that \textit{condition} was not a significant predictor, $\chi^2(1) = 1.47$, $p = .23$, nor was \textit{failure number}, $\chi^2(1) = 0.52$, $p = .47$. The interaction between \textit{failure number} and \textit{condition} approached but did not reach significance, $\chi^2(1) = 2.71$, $p = .10$. Among the covariates, \textit{previous rating} significantly predicted participants’ perceived intelligence judgement, $\chi^2(1) = 13.02$, $p < .001$. Regarding the control variables, both \textit{current failure awareness}, $\chi^2(2) = 15.53$, $p < .001$, and \textit{previous failure awareness}, $\chi^2(2) = 8.92$, $p < .05$, were significant.

Post hoc contrasts revealed a significant difference between conditions following the second failure. Participants in the heterogeneous condition reported lower perceived intelligence ratings than those in the homogeneous condition ($z = -1.97$, $p < .05$). After the third failure, however, no significant difference between conditions was observed ($z = -0.20$, $p = .84$).

\subsubsection{Effectiveness of Robot Failure Awareness}

This section reports the results of the Type~II likelihood-ratio $\chi^2$ tests conducted on Model~2, as described in Section~\ref{model3method}. The analysis investigates how participants’ perceptions of the robot, specifically trust and perceived intelligence, are influenced by the robot’s failure awareness, and whether these effects vary for different failure types. These findings address \hyperref[RQ2]{RQ2}.

\paragraph{Model-2 Results for Trust.}

\begin{figure}[t]
  \centering
  \begin{subfigure}[t]{0.48\linewidth}
    \centering
    \includegraphics[width=\linewidth]{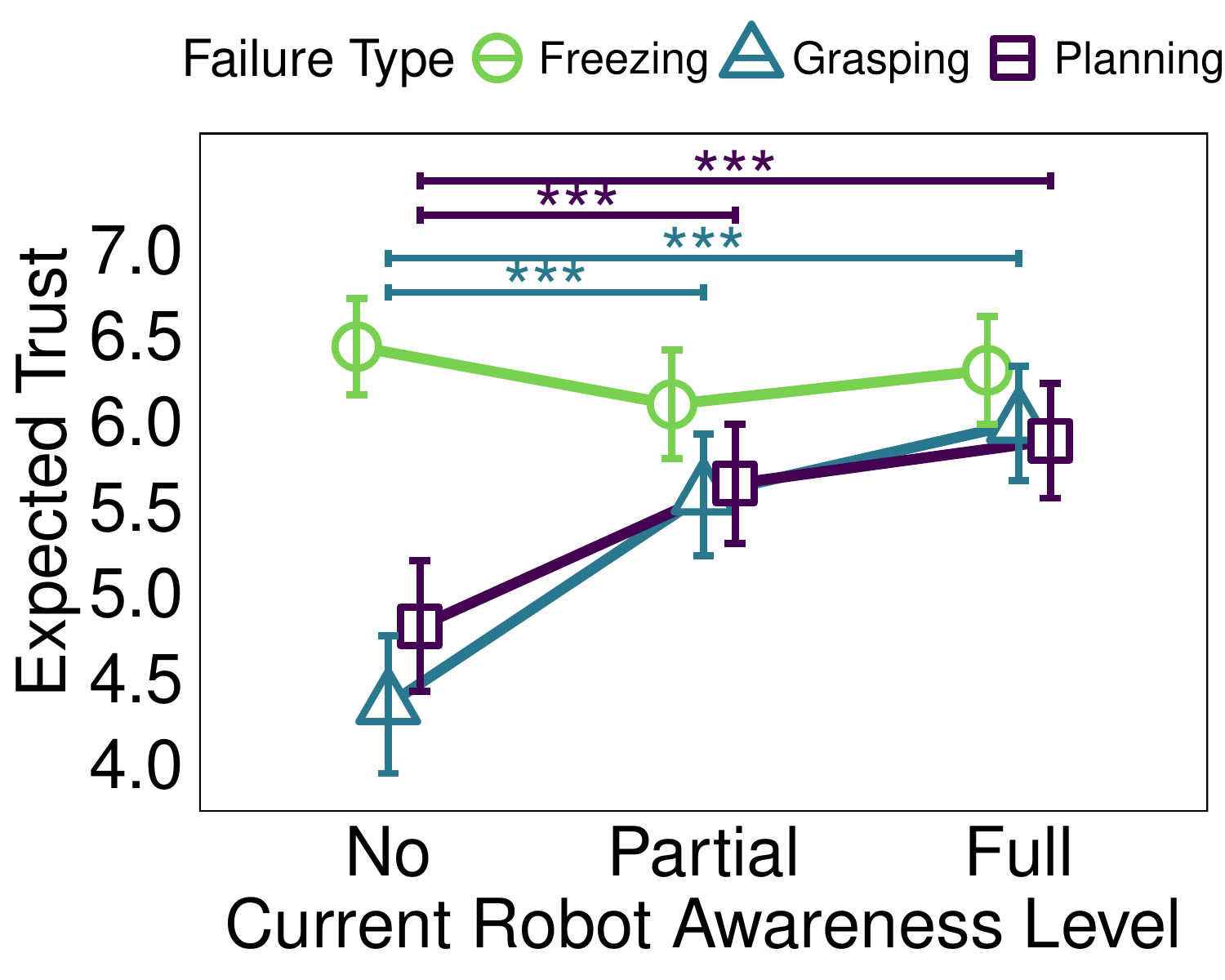}
    \caption{Trust ratings}
    \Description{The figure is a line graph showing expected trust ratings on the y-axis (ranging from 4.0 to 7.0) and current robot awareness level on the x-axis (“No Awareness,” “Partial Awareness,” and “Full Awareness”). Three lines represent different failure types: freezing, grasping, and planning. Freezing failures: trust remains high across all awareness levels, between 6.0 and 6.5. Both grasping and planning failures show a similar trend. Trust increases steadily with awareness, from about 4.5 at no awareness to over 5.5 at partial awareness and around 5.7 at full awareness. p-values above the graph indicate statistically significant differences.}
    \label{fig:trust-failureAwareness}
  \end{subfigure}
  \hfill
  \begin{subfigure}[t]{0.48\linewidth}
    \centering
    \includegraphics[width=\linewidth]{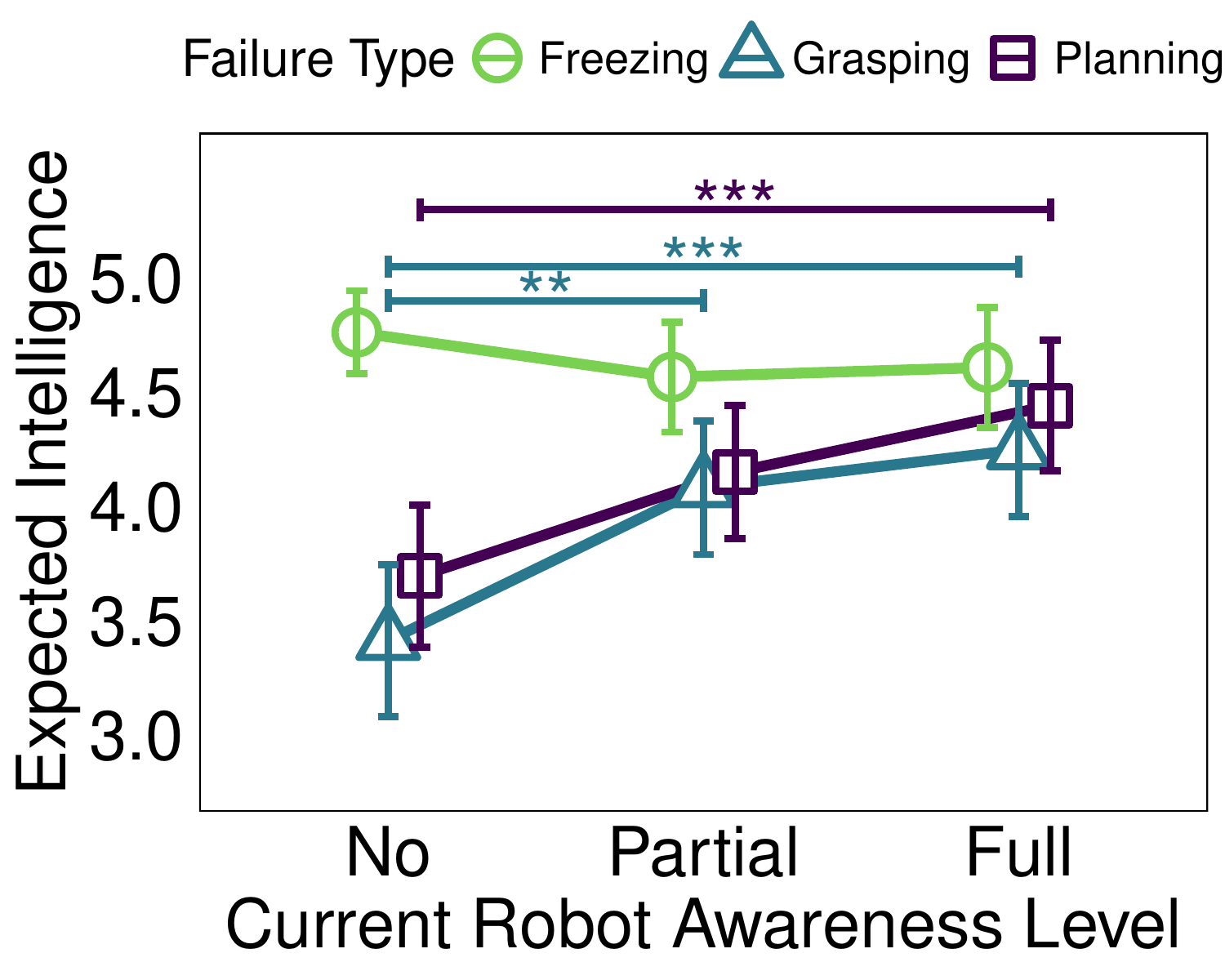}
    \caption{Perceived intelligence ratings}
    \Description{The figure is a line graph showing expected perceived intelligence ratings on the y-axis (ranging from 3.0 to 5.0) and current robot awareness level on the x-axis (“No Awareness,” “Partial Awareness,” and “Full Awareness”). Three lines represent different failure types: freezing, grasping, and planning. Freezing failures: perceived intelligence remains high across all awareness levels, between 4.5 and 5.0. Both grasping and planning failures show a similar trend. Perceived intelligence increases steadily with awareness, from about 3.5 at no awareness to over 4.0 at partial awareness and around 4.2 at full awareness. Statistical significance is indicated by asterisks, with one asterisk (*) denoting p < .05, two asterisks (**) denoting p < .01, and three asterisks (***) denoting p < .001.}
    \label{fig:intelligence-failureAwareness}
  \end{subfigure}
  \caption{Model-predicted ratings of participants’ (a) trust and (b) perceived intelligence across levels of the robot’s current failure awareness, separated by current failure type. Predictions are based on cumulative link mixed models (CLMMs) with expected ratings computed on the response scale. Error bars represent 95\% confidence intervals. Statistical significance is indicated by asterisks, with two asterisks (**) denoting p < .01, and three asterisks (***) denoting p < .001.}

  \label{fig:trust-intel-failureawareness-emms}
\end{figure}

The Type~II likelihood-ratio $\chi^2$ tests revealed significant main effects of \textit{current failure type}, $\chi^2(2) = 54.32, p < .001$, \textit{current failure awareness}, $\chi^2(2) = 40.13, p < .001$, and their interaction, $\chi^2(4) = 35.17, p < .001$. The control variables \textit{previous failure type}, $\chi^2(2) = 9.59, p < .01$, and \textit{previous failure awareness}, $\chi^2(2) = 6.64, p < .05$, were also significant. Among the covariates, \textit{pre-failure trust} was significant, $\chi^2(1) = 32.43, p < .001$, whereas \textit{failure number} was not, $\chi^2(1) = 0.13, p = .72$.

The post-hoc comparisons showed for \textit{freezing failure}, no significant differences were observed between awareness conditions (all $p > .18$). In contrast, for \textit{grasping failure}, both \textit{partial awareness} ($z = -4.86, p < .001$) and \textit{full awareness} ($z = -6.40, p < .001$) led to significantly more positive trust than \textit{no awareness}, with no difference between partial and full awareness ($p = .15$). Similarly, for \textit{planning failure}, both \textit{partial awareness} ($z = -3.61, p < .001$) and \textit{full awareness} ($z = -4.66, p < .001$) significantly outperformed \textit{no awareness}, with no difference between partial and full awareness ($p = .49$). These effects are illustrated in Figure~\ref{fig:trust-failureAwareness}, which presents the model-predicted trust ratings across failure types and levels of robot awareness.

\paragraph{Model-2 Results for Perceived Intelligence.}

The Type~II likelihood-ratio $\chi^2$ tests indicated significant main effects of \textit{current failure type}, $\chi^2(2) = 43.94, p < .001$, and \textit{current failure awareness}, $\chi^2(2) = 16.18, p < .001$, as well as their interaction, $\chi^2(4) = 17.71, p < .01$. The control variables \textit{previous failure type}, $\chi^2(2) = 6.18, p < .05$, and \textit{previous failure awareness}, $\chi^2(2) = 15.16, p < .001$, were also significant. Among the covariates, \textit{pre-failure perceived intelligence} was significant, $\chi^2(1) = 28.12, p < .001$, whereas \textit{failure number} was not, $\chi^2(1) = 0.54, p = .46$.

Post-hoc comparisons further clarified these effects. For \textit{freezing failure}, awareness level did not significantly influence participants’ perceived intelligence (all $p > .34$). For \textit{grasping failure}, however, both \textit{partial awareness} ($z = -3.17, p < .01$) and \textit{full awareness} ($z = -3.91, p < .001$) produced significantly higher perceived intelligence than \textit{no awareness}, with no significant difference between partial and full awareness ($p = .66$). In the case of \textit{planning failure}, \textit{full awareness} was associated with significantly more positive perceived ratings compared to \textit{no awareness} ($z = -3.62, p < .001$), whereas the effect of partial awareness approached significance ($p = .06$). No significant difference was observed between partial and full awareness ($p = .31$). These effects are illustrated in Figure~\ref{fig:intelligence-failureAwareness}, which shows the model-predicted intelligence ratings across failure types and levels of robot awareness.

\subsubsection{Effects of Failure Severity on Trust and Perceived Intelligence}\label{FailureHistory}

This section presents the empirical findings from Model~3, introduced in Section \ref{model2method}, which examines how users’ perceptions, specifically trust and perceived intelligence, are influenced by both the current failure severity and the severity of the preceding failure. These results address \hyperref[RQ3]{RQ3}.

To assess whether participants differentiated among the robot’s failures in terms of severity, we analysed their severity rankings for Question 1, as presented in Table \ref{TabelQual}. These rankings were converted into severity scores for each participant in the heterogeneous group (3 = highest severity, 1 = lowest severity). A Friedman test revealed a significant difference in perceived severity across the three failure types, $\chi^2(2) = 21.48$, $p < .001$. Follow-up pairwise Wilcoxon signed-rank tests with Holm correction showed that both the planning failure and the grasping failure were rated as significantly more severe than the freezing failure ($p < .001$ and $p < .001$, respectively). No significant difference was found between the planning and grasping failures ($p = .66$), indicating that participants considered these two failure types similarly severe. Accordingly, for subsequent analyses, the planning and grasping failures were grouped as high-severity failures, whereas the freezing failure was treated as a low-severity failure.

\begin{figure}[t]
  \centering
  \begin{subfigure}[t]{0.48\linewidth}
    \centering
    \includegraphics[width=\linewidth]{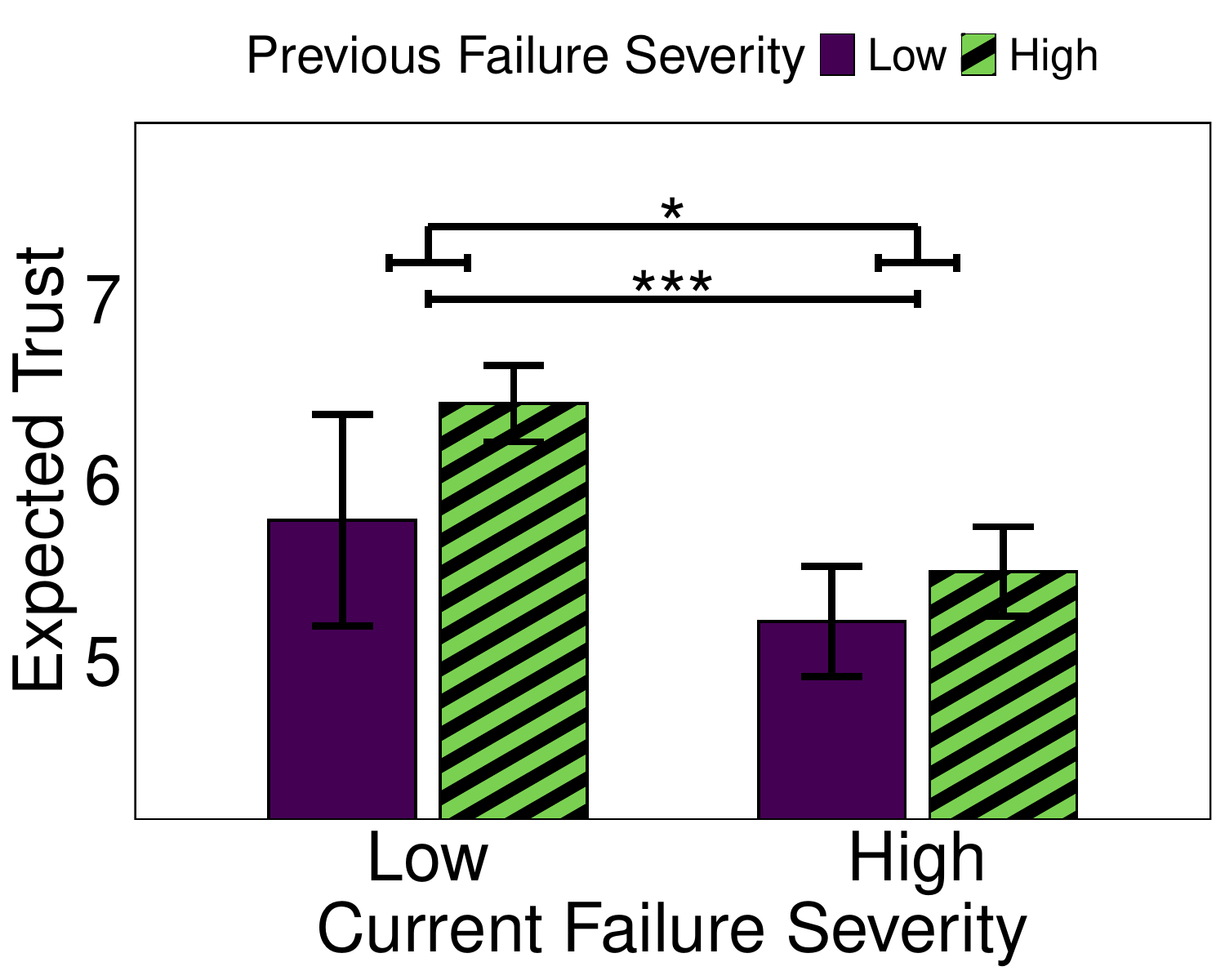}
    \caption{Trust ratings}
    \Description{The figure is a bar chart showing expected trust ratings on the y-axis (ranging from 5.0 to 7.0) and current failure severity on the x-axis (from low to high). For each level of current failure severity, two bars are shown: a dark bar representing cases where the previous failure severity was low, and a light bar representing cases where the previous failure severity was high. The chart shows that trust decreases as current failure severity increases, with trust consistently lower when the previous failure severity was high. p-values above the bars indicate statistically significant differences between the conditions.}

    \label{fig:trust-failureType}
  \end{subfigure}
  \hfill
  \begin{subfigure}[t]{0.48\linewidth}
    \centering
    \includegraphics[width=\linewidth]{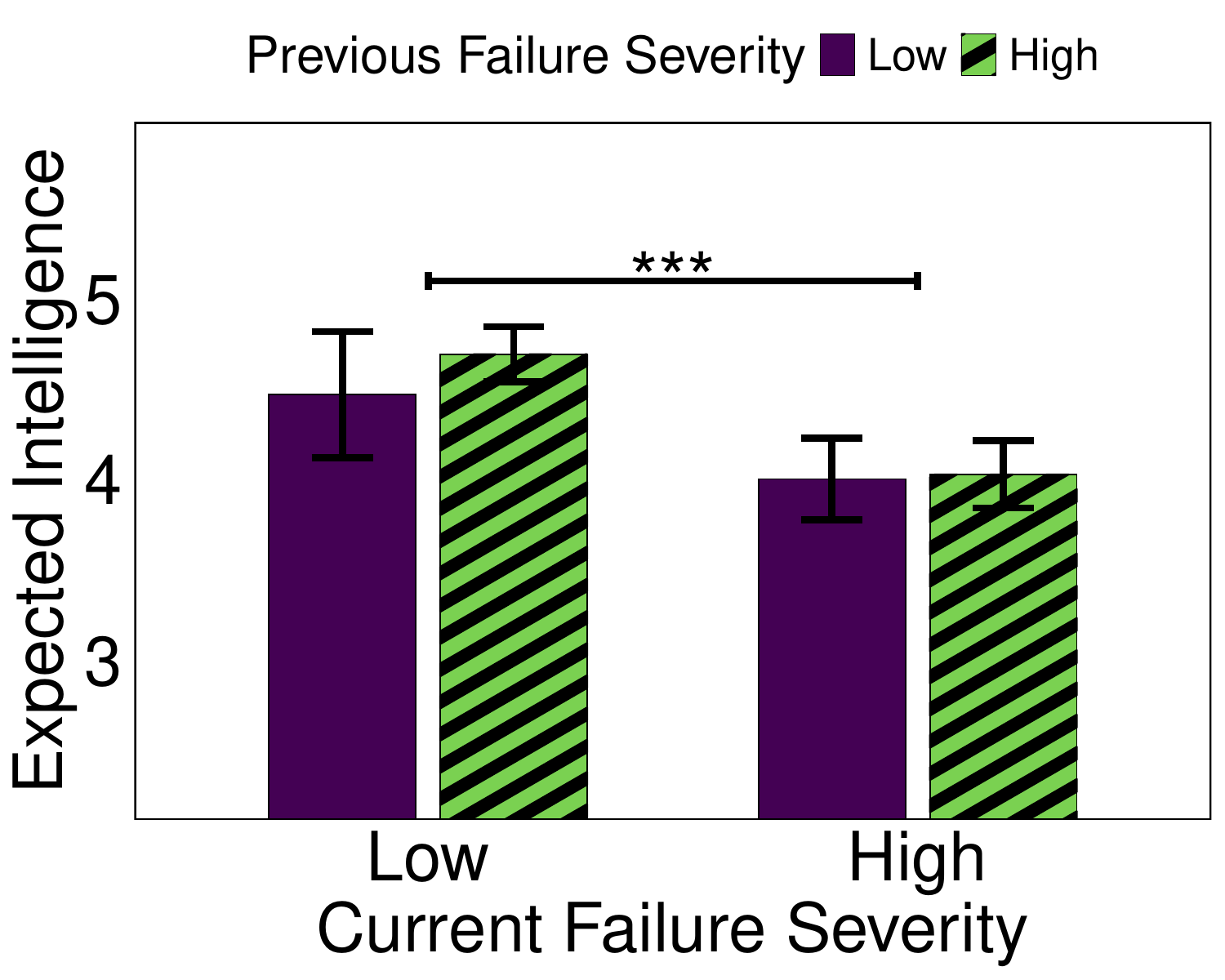}
    \caption{Perceived intelligence ratings}
    \Description{The figure is a bar chart showing expected perceived intelligence ratings on the y-axis (ranging from 3.0 to 5.0) and current failure severity on the x-axis (from low to high). For each level of current failure severity, two bars are shown: a dark bar representing cases where the previous failure severity was low, and a light bar representing cases where the previous failure severity was high. The chart shows that perceived intelligence decreases as current failure severity increases. Asterisks above the bars indicate statistically significant differences between the conditions.}

    \label{fig:intelligence-failureType}
  \end{subfigure}
  \caption{Expected scores (model predictions) of participants’ trust (left) and perceived intelligence (right) across current failure severity, separated by previous failure severity.  Error bars represent 95\% confidence intervals. Statistical significance is indicated by asterisks, with one asterisk (*) denoting p < .05, and three asterisks (***) denoting p < .001.}
  \label{fig:trust-intel-emms}
\end{figure}

\paragraph{Model-3 Results for Trust}

To evaluate the influence of predictors on participants’ trust ratings, Type~II likelihood–ratio $\chi^2$ tests were conducted. The analysis revealed significant effects of the \textit{severity of the current failure}, $\chi^2(1) = 53.27$, $p < .001$, and the \textit{severity of the preceding failure}, $\chi^2(1) = 5.93$, $p < .05$. The interaction between the severity of current and preceding failures was not significant, $\chi^2(1) = 1.70$, $p = .19$. Among the covariates, the \textit{pre-failure trust rating} was significant, $\chi^2(1) = 23.16$, $p < .001$, whereas the \textit{failure number} was not, $\chi^2(1) = 0.02$, $p = .90$. For the control variables, both \textit{current failure awareness} and \textit{preceding failure awareness} were significant, $\chi^2(2) = 40.54$, $p < .001$, and $\chi^2(2) = 7.62$, $p < .05$, respectively. The expected trust ratings for \textit{current failure severity} by \textit{preceding failure severity} are visualised in Figure~\ref{fig:trust-failureType}.

\paragraph{Model-3 Results for Perceived Intelligence.}

Similar to the analysis of trust ratings, Type~II likelihood–ratio $\chi^2$ tests were conducted. The analysis revealed significant effects of the \textit{severity of the current failure}, $\chi^2(1) = 39.63$, $p < .001$. The \textit{severity of the preceding failure}, $\chi^2(1) = 0.54$, $p = .46$, and the interaction between the severity of current and preceding failures were not significant, $\chi^2(1) = 1.07$, $p = .30$. Among the covariates, the \textit{pre-failure intelligence rating} was significant, $\chi^2(1) = 24.15$, $p < .001$, whereas the \textit{failure number} was not, $\chi^2(1) = 0.28$, $p = .60$. For the control variables, both \textit{current failure awareness} and \textit{preceding failure awareness} were significant, $\chi^2(2) = 15.86$, $p < .001$, and $\chi^2(2) = 13.22$, $p < .01$, respectively. The expected intelligence ratings for \textit{current failure severity} by \textit{preceding failure severity} are visualised in Figure~\ref{fig:intelligence-failureType}.

\subsection{Qualitative Results}\label{Qualitative Results}

This section analyses participants’ open-ended comments, with two qualitative objectives. Section \ref{Qual1} and \ref{Qual2} explore the rationales and emotions that help explain the trust dynamics addressed in \hyperref[RQ2]{RQ2} and \hyperref[RQ3]{RQ3}.

\subsubsection{Perceived Severity of Failures}\label{Qual1}

The first open-ended question asked participants to evaluate which robot failure they perceived as most and least severe, and to explain the reasoning behind their rankings. A reflexive thematic analysis was carried out to explore participants’ reasoning behind their severity judgements for each failure type.


\paragraph{Grasping failure.} Participants who rated grasping failure as the most severe often associated it with \textbf{perceived incompetence and unreliability}, expressing concerns about the robot’s fundamental abilities. For instance, one participant noted, ``it [grasping failure] made the robot look like it was broken'' (P12), while another remarked, ``the size of the grip of each piece should be around the same, and the robot was already trained to pick up similar pieces ahead'' (P41). Some also highlighted a \textbf{violation of expected human-like behaviour}, describing the robot’s action as unnatural and disruptive to collaborative flow. As one participant explained, ``this [grasping failure] does not (rarely) happen for humans, like picking air instead of something concrete'' (P26).

Conversely, some participants regarded grasping failure as the least severe due to their \textbf{easy recognisability}, referring to how clearly observable the error appeared during the task. Others attributed this judgment to the robot’s \textbf{mechanical limitations}, recognising the failure as a plausible outcome of hardware constraints rather than a flaw in decision-making. For instance, P38 noted, ``failing to pick up an object could be passed off as a mechanical issue … but [it] would have made me more annoyed with the software.''

\paragraph{Planning failure.} Participants who perceived \textit{Planning} failure as the most severe often cited the \textbf{high-stakes consequences}—defined as the potential for the error to affect subsequent actions. One participant commented, ``if a robot places something where it isn't meant to be, it can cause more problems for the later process'' (P32). Others described the \textbf{cognitive confusion} these errors caused, referring to the mental uncertainty and difficulty in attributing fault, as they struggled to determine whether the robot or they themselves were responsible for the mistake. For instance, P6 noted, ``it confused me—it made me believe I was analysing the puzzle wrong.''

Those who rated \textit{Planning} failure as the least severe typically emphasised that such errors were \textbf{easily correctable}, meaning they could be resolved without the need for expert assistance. Additionally, some participants interpreted these mistakes as \textbf{relatable human-like errors}, which elicited empathy rather than frustration. For instance, one participant noted, ``that was because I could easily correct it myself'' (P14), while another remarked, ``this happens sometimes for humans'' (P26).

Across both \textit{Grasping} and \textit{Planning} failures, participants’ severity judgements were strongly influenced by the robot’s level of awareness and its response to the error. Failures were generally perceived as more severe when the robot \textbf{lacked awareness or failed to correct the mistake}. Conversely, the same types of failures were mostly regarded as least severe when the robot exhibited \textbf{immediate awareness and initiated corrective action}. For instance, in the case of grasping failure, P2 remarked, ``I think failing to pick up an object was the most severe because he didn't give me a reply,'' whereas P6 noted, ``the least severe was failing to pick up the object as it proceeded to do so right away after recognising its mistake.''

\paragraph{Freezing failure.}  The dominant theme among participants who rated freezing failure as most severe was the \textbf{disruption of task flow and user frustration}. The unexpected pause in the robot's actions was perceived as confusing and disengaging. As P50 noted, ``it created an uncomfortable awkwardness as if I was waiting for it to do something but it wouldn't.''

However, many participants who rated \textit{Freezing} failure as the least severe emphasised their \textbf{low impact and lack of serious consequences}. This theme reflects the perception that temporary pauses did not hinder task completion or require user intervention. As P24 explained, ``the robot just needed more time; it finished the task.'' Others viewed the \textit{Freezing} behaviour as \textbf{systemic or expected}, interpreting it not as a malfunction but as a natural processing delay. For instance, P47 remarked, ``it's more like a `processing time'.''

\subsubsection{Perceptions of Robot Recovery Strategies}\label{Qual2}

We conducted a reflexive thematic analysis to explore participants’ perceptions of the robot’s recovery strategies, focusing on how different levels of the robot’s awareness of its own failures influenced user experience. Our aim was to understand the subjective impact of these responses and to identify the underlying reasons, expressed as thematic constructs.

When the robot exhibited \textbf{no awareness} of its failure—failing to offer any verbal or physical response—participants overwhelmingly described the experience as negative. A dominant theme that emerged was \textbf{Cognitive Uncertainty and Confusion}, referring to the ambiguity and indecision participants experienced regarding their role and next steps in the task. As P35 noted, ``I was very lost and unsure whether to help or not,'' and P2 remarked, ``I didn't know what I should do next.'' Another commonly identified theme was \textbf{Perceived Incompetence}, in which the robot’s silence was interpreted as a lack of intelligence or reliability. P15 commented, ``when it wasn't aware of its mistake, it seemed less reliable,'' while P12 reflected, ``the fact that the robot did not react or comment on its mistake made it look like it could not think.'' A third theme, \textbf{Frustration}, captured emotional responses to the robot’s lack of engagement, with P40 stating, ``the instance where it didn't make a reaction annoyed me.''

Conversely, when the robot demonstrated \textbf{full awareness} of its failure by acknowledging the mistake and initiating corrective action, participants reported predominantly positive experiences. One primary theme was \textbf{Autonomous Competence}, reflecting admiration for the robot’s ability to independently identify and correct its errors. P35 remarked, ``its awareness of the mistake, and his ability to quickly fix the mistake impressed me,'' while P53 stated, ``after trying to correct the mistakes I felt reliability.'' Another theme, \textbf{Explanatory Apology}, captures how the robot’s combination of explanation and apology reinforced participants’ confidence in the robot. For instance, P37 shared, ``I felt more pleased when the robot indicated the reason for its mistake and apologised,'' and P14 described, ``I found the robot's reaction when it failed very charming. It admitted its mistake and said that it would try again.'' Finally, some participants highlighted \textbf{Human-Like Social Presence}, perceiving the robot’s responsive behaviour as more natural and relatable. As P34 noted, ``I felt it was more `lively' and human-like.''

An analysis of participants’ responses to the robot’s \textbf{partial awareness} of its failures revealed both positive and negative perceptions. Some participants appreciated the robot’s behaviour as providing \textbf{actionable guidance}, whereby the robot's acknowledgement of the issue and request for assistance helped clarify the task. For example, P35 remarked, ``the robot was aware and gave me instructions to correct its mistake; that was very clear for me and made it simple.'' Others, however, viewed the same behaviour as a \textbf{delegation of agency}, interpreting the robot’s request as an inappropriate transfer of responsibility. As P26 noted, ``it asked me to help it fix [the mistake], rather than noticing and trying to fix it itself.''

Similarly, some participants described the robot’s partial awareness as fostering \textbf{collaborative engagement}, highlighting a sense of joint problem-solving and shared effort. P31 explained, ``when the robot acknowledged its mistake and asked for my help, I was willing to help fix it and more sympathetic towards it.'' In contrast, others interpreted this behaviour as a \textbf{competency violation}, expressing concerns about the robot’s intelligence and autonomy. As P32 reflected, ``when it recognised the mistake but did not fix the issue itself, that made me a little troubled because that means the robot is incapable of amending this problem without help.''



\section{Discussion}

This study examined how three factors shape user evaluations across repeated robot failures: the sequence of failure types (homogeneous vs. heterogeneous), perceived severity (current vs. previous failures), and robot awareness (none, partial, full) of its own errors. Our objective is to translate these findings into sequence-sensitive design principles that help preserve user trust and perceived intelligence in the face of repeated failures. The following subsections outline each factor in turn and discuss their respective practical implications.



\subsection{Beyond Frequency: Mixed Failure Sequences Erode Trust and Perceived Intelligence}

Our results indicate that user evaluations are shaped by the type of failures and their sequence. Their order and diversity influence how people judge the robot. Although trust did not differ significantly between conditions, the overall pattern shows that heterogeneous sequences tended to produce larger declines than homogeneous ones. For perceived intelligence, this effect was statistically evident: participants rated the robot as less intelligent after a heterogeneous sequence, particularly at the second failure. This suggests that users are sensitive to the diversity of problems a robot displays, and that varying failure types may be interpreted as signs of broader system unreliability.

When different kinds of failures occur in sequence, participants may infer that the robot’s issues are widespread rather than localised, reducing both trust and perceived intelligence. Conversely, when the same failure repeats, users can attribute it to a specific subsystem, making it appear more contained. Repeated failures may also make the robot’s behaviour somewhat more predictable—users know what to expect and how the robot is likely to respond. A shift to a different failure type, however, disrupts these emerging expectations, adds an element of surprise, and makes the robot seem less stable overall. This interpretation aligns with EVT, which predicts that unexpected or inconsistent events trigger stronger negative reactions than anticipated ones. In our study, the second failure appears particularly important: when it differed from the first, it violated users’ expectations, leading to sharper declines in perceived intelligence. Even though trust did not show statistically significant differences, the descriptive pattern followed the same direction.

These findings yield several implications for designing and deploying adaptive failure-recovery strategies. First, systems should consider the sequence of failures rather than treating each one in isolation. When a failure differs from the preceding one, repair strategies may need to escalate—providing more detailed explanations, running broader diagnostic checks, or adopting more conservative actions to minimise the chance of additional errors. Second, designers should monitor the diversity of failure types, not only their frequency. An expanding range of failures within a single interaction may signal to users that the system is unstable, warranting stronger verification mechanisms or dynamic adjustments to maintain user confidence. This study therefore extends existing trust calibration models proposed by Nesset et al.~\cite{nesset_robot_2023} and Esterwood et al.~\cite{esterwood_having_2022} by demonstrating that evaluations of trust and perceived intelligence depend not just on failure frequency but on the type and temporal structure of failures.

\subsection{Obliviousness costs: silence on noticeable errors lowers trust and perceived intelligence.}

The results of model~2 and the thematic analysis jointly indicate that the robot’s \emph{obliviousness} to failure was generally associated with poorer user experience for grasping and planning errors. When the robot offered no acknowledgement, participants reported lower trust and perceived intelligence alongside \textit{cognitive uncertainty}, \textit{perceived incompetence}, and \textit{frustration} (e.g., ``I didn't know what I should do next,'' P2; ``it seemed less reliable,'' P15). Importantly, this pattern did \emph{not} extend to the freezing failure, for which awareness level showed no reliable influence on evaluations. This study extends existing trust calibration models by comparing different robot reactions after a failure in terms of their level of awareness, rather than evaluating individual approaches (e.g., apology or explanation). It also demonstrates how the amount of awareness required varies depending on the type of failure.

Awareness effects are not uniform across failure types. For high-severity or noticeable errors (grasping, planning), both \textit{partial}  and \textit{full}  awareness reliably improved trust compared to \textit{no awareness}, with no reliable difference between the two levels. Perceived intelligence followed the same pattern for grasping and showed a weaker but consistent trend for planning, where full awareness significantly outperformed no awareness and partial awareness moved in the same direction. These results suggest that in manipulation tasks where the problem and its remedy are interpretable to users, \emph{acknowledgement}, paired with either autonomous correction or a request for help, provides most of the benefit. 

By contrast, freezing failure showed no reliable benefit of awareness on either trust or perceived intelligence. One interpretation is that awareness is most effective when paired with an immediate and necessary remedy. When no physical fixing is required for task completion, as in freezing, additional communicative acts may not yield higher evaluations. Another possibility is that explicitly labelling freezing as a failure may unintentionally undermine trust, while users might otherwise tolerate temporary freezing.
Indeed, such short stalls are common in autonomous systems, frequently arising from perception latencies (e.g., processing algorithms \cite{liu_prophet_2022}, sensor noise \cite{falanga_how_2019}) or collision-avoidance planning \cite{trautman_unfreezing_2010}. Because these delays are routine, users may already normalise them, which can limit, or even backfire on, the perceived value of explicit awareness.

The qualitative results also highlight a \emph{trade-off} inherent to \textit{partial awareness}. Some participants experienced the robot’s request for help as \textit{actionable guidance} and \textit{collaborative engagement} (``I was willing to help fix it'' P31), while others interpreted the same cue as a \textit{delegation of agency} or a \textit{competency violation} (“rather than … trying to fix it itself'' P26; ``incapable of amending this problem without help'' P32). By contrast, the quantitative results show that partial awareness performs nearly as well as full awareness in manipulation failures, with only a slight advantage for full awareness in terms of trust and perceived intelligence. These findings suggest that although users interpret partial awareness in divergent ways, even incomplete explanations of failure can sustain trust in HRI, as the communicative act itself signals transparency and accountability.


The findings point to important design implications for awareness strategies in human–robot interaction. Failure detection can be achieved either through the robot’s internal monitoring mechanisms or through observation of users’ reactions to the failure. Depending on the operational environment and the types of potential failures, designers must determine the level of awareness a robot should demonstrate and the type of reaction required to sufficiently regain user trust. In some contexts, it may not be necessary for the robot to detect and explicitly communicate that a failure has occurred. This was evident in our design, where awareness cues during freezing failures reduced both trust and perceived intelligence because the robot stated that a failure had occurred, even though the failure itself was subtle and largely unnoticed by users. In other contexts, however, the robot does need to detect that a failure occurred and identify the specific failure type. In our experiment, for planning and grasping failures, which were noticeable to users, combining acknowledgement with either autonomous correction or a minimal request for help was required to stabilise trust; full self-correction was preferred but not strictly necessary.

\subsection{Expectancy violation: severe after mild amplifies trust loss.}

Our findings demonstrate that the perceived severity of the current failure is the primary determinant of user evaluations: higher severity reliably reduces both trust and perceived intelligence. In our study, grasping and planning errors, classified as high-severity, were consistently rated as more severe, with participants describing them as ``incompetent,'' ``unnatural,'' or ``high-stakes''. By contrast, the freezing error, classified as low-severity, was often dismissed as a simple “processing” pause. These findings align with prior work showing that failure severity shapes user reactions \cite{stiber_not_2021} and that different types of navigation errors can lead to varying degrees of trust reduction \cite{garcia_goo_being_2025}.


Trust appeared to be a history-sensitive construct, as the severity of the previous failure affected trust on the subsequent trial, whereas perceived intelligence showed little evidence of such carry-over effects. This temporal sensitivity aligns with EVT \cite{burgoon_interpersonal_1993, burgoon_nonverbal_1988}, which proposes that people react more strongly when events deviate from their emerging expectations. In our study, when a failure occurred after a low-severity one, trust declined more sharply than when it followed a severe failure, indicating that a severe error violates a user’s expectation of relative stability established by the prior mild failure. By contrast, once a severe failure has already occurred, a subsequent failure, while still detrimental, elicits less additional shock because the expectation of unreliability has already been set. Consequently, trust is more negatively impacted when a severe failure follows a low-severity one. This pattern reflects a sequence-dependent expectancy violation and also extends current trust calibration models by showing that trust is shaped not only by failure magnitude but also by the order in which failures occur. Interestingly, this sequencing effect did not appear for perceived intelligence, which seemed to track only the immediate performance signal rather than incorporating prior failures. This dissociation suggests that trust integrates experiences over time, whereas perceived intelligence functions more as a momentary evaluation of current competence. Prior psychological research supports this interpretation, showing that severe events following milder ones heighten expectancy violations and amplify negative perceptions \cite{baumeister_bad_2001}.

These findings highlight several implications for the design and deployment of interactive robots. Severe failures should be handled much like alarms, requiring clear acknowledgement and prompt corrective action to limit damage to both trust and perceived intelligence. Importantly, trust-repair strategies should also take the sequence of failures into account. For example, when a severe failure occurs after a low-severity one, a pattern that tends to produce sharper trust declines, the system should escalate its repair efforts. This may involve offering richer explanations, or introducing additional safety checks, thereby helping to stabilise trust in more vulnerable situations.

\subsection{Limitations and Paths Forward}

This study was conducted in a controlled laboratory environment using a Tangram puzzle task and the TIAGo robot, which enabled precise manipulation of failure types and awareness levels. However, such a setup may not fully generalise to real-world scenarios in which task complexity, environmental unpredictability, and safety stakes are substantially higher. In addition, the robot’s awareness behaviours were pre-scripted and implemented through a combination of verbal explanations and predefined physical repair actions, which may not fully capture the flexibility and adaptivity of real-time autonomous responses. Only three failure instances were included per session, limiting insight into longer-term trust dynamics across extended human–robot interactions. Furthermore, the sample size was relatively moderate (N = 54); larger and more diverse participant samples would strengthen the robustness and generalisability of the observed effects. Due to ethical constraints, all failures were required to be safe, non-damaging, and free of safety risks, which restricted our ability to manipulate failure severity more explicitly. Finally, the freezing failure consistently resulted in a successful task outcome, distinguishing it from the grasping and planning failures in terms of perceived task impact.

Building on the present findings, future research should examine robot failures and trust repair in more ecologically valid contexts such as domestic, industrial, or healthcare settings, where task complexity and safety stakes are higher. Extending beyond pre-scripted behaviours, adaptive repair strategies that leverage real-time user state detection (e.g., stress, gaze, hesitation) could provide more flexible and context-sensitive trust recovery. Longer-term studies with more frequent or varied failures are also needed to capture how trust stabilises, erodes, or recovers across extended collaborations. In addition, investigating failures of greater severity or consequence would provide insight into how well different trust repair strategies hold up under higher-risk conditions. Finally, moving from dyadic interactions to multi-human or multi-robot team contexts could reveal how failure awareness and sequence effects shape group trust and coordination, informing the development of design frameworks for scalable, adaptive trust repair in collaborative robotics.


\section{Conclusion}


This work advances our understanding of how failure sequence, robot awareness, and failure severity jointly shape user trust and perceptions of intelligence in collaborative interaction. Our findings show that both trust and perceived intelligence are history-sensitive: heterogeneous sequences of failures reduce perceived intelligence more than homogeneous ones, and prior severe failures exert lasting negative effects on trust over time. Robot awareness further moderates these dynamics, with acknowledgement and repair stabilising trust in noticeable failures (e.g., grasping and planning) but offering no benefit, or even backfiring, for less interpretable errors such as freezing. These insights highlight the need for sequence-aware and context-sensitive trust repair strategies, where robots escalate repair efforts when severe failures follow mild ones, treat diverse failures as signals of broader instability, and adjust their explanations depending on the interpretability of the error.




\begin{acks}
This work is partly supported by the Australian Research Council (Grant No: FT250100459 and  CE260100108)
\end{acks}

\bibliographystyle{ACM-Reference-Format}
\bibliography{references,sample-base}

\newpage
\onecolumn

\appendix 
\renewcommand{\thetable}{A\arabic{table}}
\setcounter{table}{0}  
\section{Appendix: Model Results}

Tables~\ref{tab:A1}--\ref{tab:A3} present the detailed results for the three cumulative link mixed models, corresponding respectively to Formulas~\ref{eq:model1}, \ref{eq:model2}, and \ref{eq:model3}.

\begin{equation}
\begin{aligned}
\text{Dependent\_Variable} \sim\;&
\text{Prefailure\_Rating}
+ \text{Sequence\_Condition} \times \text{Failure\_Number} + \text{Current\_Awareness} + 
\\
& \text{Previous\_Awareness}
+ (1 \mid \text{Participant})
+ (1 \mid \text{Item})
\end{aligned}
\label{eq:model1}
\end{equation}

\begin{equation}
\begin{aligned}
\text{Dependent\_Variable} \sim\;&
\text{Prefailure\_Rating}
+ \text{Failure\_Number}
+ \text{Current\_Failure\_Type} \times \text{Current\_Awareness} + \\
& \text{Previous\_Awareness}
+ (1 \mid \text{Participant})
+ (1 \mid \text{Item})
\end{aligned}
\label{eq:model2}
\end{equation}

\begin{equation}
\begin{aligned}
\text{Dependent\_Variable} \sim\;&
\text{Prefailure\_Rating}
+ \text{Failure\_Number}
+ \text{Current\_Awareness} + \text{Previous\_Awareness}+\\
&
 \text{Severity\_Current\_Failure} \times 
   \text{Severity\_Previous\_Failure} + (1 \mid \text{Participant})
+ (1 \mid \text{Item})
\end{aligned}
\label{eq:model3}
\end{equation}

Where \text{Dependent\_Variable} represents the rating from one of the scales (trust or perceived intelligence);
\text{Prefailure\_Rating} reflects the rating given by the participant on the preceding failure;
\text{Failure\_Number} indicates the position of the current failure within the sequence;
\text{Current\_Awareness} and \text{Previous\_Awareness} denote the robot’s verbal awareness level in response to failure (no, partial, or full awareness) on the current and previous trial, respectively;
\text{Severity\_Current\_Failure} and \text{Severity\_Previous\_Failure} represent the severity level of the current and preceding failures;
\text{Sequence\_Condition} specifies the between-sequence manipulation (homogeneous vs. heterogeneous failure sequence);
\text{Current\_Failure\_Type} identifies the category of the current failure;
and the interaction terms assess whether the effect of one predictor depends on the level of another.
Finally, (\text{Participant}) and (\text{Item}) indicate random intercepts, allowing baseline ratings to vary across participants and across scale items.

\begin{table}[h!]
\centering
\caption{This table presents the results of the CLMM analysis for Model 1.}
\label{tab:A1}
\Description{This table reports the results of the cumulative link mixed model (CLMM) analysis for Model 1, examining the effects of failure number and condition (homogeneous vs.\ heterogeneous) on participants’ trust and perceived intelligence ratings. For each outcome variable, the table presents estimated coefficients, standard errors, z-values, and p-values for the main predictors and control variables, including previous rating and current and previous awareness levels. Random effects for participants and items are also reported, along with model fit indices (log-likelihood and AIC) for both trust and perceived intelligence models.}

\renewcommand{\arraystretch}{1.2}
\setlength{\tabcolsep}{6pt}

\begin{tabular}{
    >{\centering\arraybackslash}m{4.2cm} |
    >{\centering\arraybackslash}m{0.93cm} 
    >{\centering\arraybackslash}m{0.8cm} 
    >{\centering\arraybackslash}m{0.8cm} 
    >{\centering\arraybackslash}m{0.8cm} |
    >{\centering\arraybackslash}m{0.93cm} 
    >{\centering\arraybackslash}m{0.8cm} 
    >{\centering\arraybackslash}m{0.8cm} 
    >{\centering\arraybackslash}m{0.8cm} 
}
\toprule
\textbf{Model 1} 
  & \multicolumn{4}{c|}{\textbf{Trust}} 
  & \multicolumn{4}{c}{\textbf{Perceived Intelligence}} \\ 
\cmidrule(lr){2-9}
\textbf{Fixed Effects} 
  & \textbf{Est} & \textbf{SE} & \textbf{z} & \textbf{p}
  & \textbf{Est} & \textbf{SE} & \textbf{z} & \textbf{p} \\ 
\midrule

\multicolumn{9}{l}{\textbf{Main Predictors}} \\

Condition (Homogeneous)
  & 0.466 & 0.546 & 0.853 & .394
  & 0.913 & 0.481 & 1.900 & .057 \\

Failure Number (Third)
  & -0.071 & 0.229 & -0.309 & .758
  & 0.103 & 0.284 & 0.363 & .717 \\

Failure Number (Third) × Condition (Homogeneous)
  & 0.014 & 0.405 & 0.034 & .973
  & -0.821 & 0.500 & -1.642 & .101 \\

\midrule

\multicolumn{9}{l}{\textbf{Covariate/Control Factors}} \\

Previous Rating
  & 0.296 & 0.136 & 2.171 & <.05
  & 0.724 & 0.194 & 3.733 & <.001 \\

Current Awareness (Partial) 
  & 1.007 & 0.297 & 3.395 & <.001
  & 0.828 & 0.349 & 2.376 & <.05 \\

Current Awareness (Full) 
  & 1.718 & 0.310 & 5.536 & <.001
  & 1.434 & 0.377 & 3.808 & <.001 \\

Previous Awareness (Partial) 
  & 0.845 & 0.310 & 2.725 & <.01
  & 0.191 & 0.377 & 0.507 & .612 \\

Previous Awareness (Full) 
  & 0.397 & 0.311 & 1.276 & .202
  & -0.842 & 0.387 & -2.177 & <.05 \\

\midrule

\textbf{Random Effects} 
  & \textbf{Var} & \textbf{SD} &  &  
  & \textbf{Var} & \textbf{SD} &  &  \\ 
\midrule
Participant 
  & 2.616 & 1.617 &  &  
  & 1.233 & 1.110 &  &  \\
Item 
  & 0.000 & 0.000 &  &  
  & 0.030 & 0.174 &  &  \\
\midrule

\textbf{Model Fit} 
   & \textbf{logLik} & \textbf{AIC} &  &
   & \textbf{logLik} & \textbf{AIC} &  & \\ 
\midrule
   & -557.09 & 1146.19 &  &
   & -324.62 & 677.24 &  &  \\
\bottomrule
\end{tabular}
\end{table}

\begin{table}[h!]
\centering
\caption{This table presents the results of the CLMM analysis for Model 2.}
\Description{This table presents the results of the cumulative link mixed model (CLMM) analysis for Model 2, which examines how different failure types and levels of robot awareness predict trust and perceived intelligence ratings. The table reports estimated coefficients, standard errors, z-values, and p-values for main predictors, including current failure type and current awareness, as well as their interaction terms. Covariates such as previous rating, failure number, previous failure type, and previous awareness are also included. Random effects for participants and items are shown, along with model fit statistics (log-likelihood and AIC) for both the trust and perceived intelligence models.}

\label{tab:A2}
\renewcommand{\arraystretch}{1.2}
\setlength{\tabcolsep}{6pt}

\begin{tabular}{
    >{\centering\arraybackslash}m{4.2cm} |
    >{\centering\arraybackslash}m{0.93cm} 
    >{\centering\arraybackslash}m{0.8cm} 
    >{\centering\arraybackslash}m{0.8cm} 
    >{\centering\arraybackslash}m{0.8cm} |
    >{\centering\arraybackslash}m{0.93cm} 
    >{\centering\arraybackslash}m{0.8cm} 
    >{\centering\arraybackslash}m{0.8cm} 
    >{\centering\arraybackslash}m{0.8cm} 
}
\toprule
\textbf{Model 2} 
  & \multicolumn{4}{c|}{\textbf{Trust}} 
  & \multicolumn{4}{c}{\textbf{Perceived Intelligence}} \\ 
\cmidrule(lr){2-9}
\textbf{Fixed Effects} 
  & \textbf{Est} & \textbf{SE} & \textbf{z} & \textbf{p}
  & \textbf{Est} & \textbf{SE} & \textbf{z} & \textbf{p} \\ 
\midrule

\multicolumn{9}{l}{\textbf{Main Predictors}} \\

Current Failure Type (Grasping)
  & -4.990 & 0.627 & -7.954 & <.001
  & -4.621 & 0.773 & -5.979 & <.001 \\

Current Failure Type (Planning)
  & -4.009 & 0.590 & -6.789 & <.001
  & -3.811 & 0.709 & -5.379 & <.001 \\

Current Awareness (Partial) 
  & -0.967 & 0.546 & -1.771 & .076
  & -0.928 & 0.665 & -1.396 & .163 \\

Current Awareness (Full) 
  & -0.415 & 0.564 & -0.736 & .462
  & -0.755 & 0.723 & -1.043 & .297 \\

Current Failure (Grasping) × Current Awareness  (Partial)
  & 3.669 & 0.787 & 4.663 & <.001
  & 2.936 & 0.924 & 3.176 & <.01 \\

Current Failure (Planning) × Current Awareness  (Partial)
  & 2.845 & 0.766 & 3.712 & <.001
  & 2.347 & 0.921 & 2.548 & <.05 \\

Current Failure (Grasping) × Current Awareness  (Full)
  & 4.149 & 0.821 & 5.054 & <.001
  & 3.304 & 0.967 & 3.417 & <.001 \\

Current Failure (Planning) × Current Awareness  (Full)
  & 2.909 & 0.778 & 3.739 & <.001
  & 3.159 & 0.982 & 3.219 & <.01 \\

\midrule

\multicolumn{9}{l}{\textbf{Covariate/Control Factors}} \\

Previous Rating
  & 0.881 & 0.153 & 5.765 & <.001
  & 1.158 & 0.213 & 5.424 & <.001 \\

Failure Number (Third)
  & -0.070 & 0.194 & -0.362 & .717
  & -0.179 & 0.245 & -0.733 & .464 \\

Previous Failure Type (Grasping)
  & 0.841 & 0.356 & 2.358 & <.05
  & -0.172 & 0.393 & -0.436 & .663 \\

Previous Failure Type (Planning)
  & 1.067 & 0.350 & 3.052 & <.01
  & 0.748 & 0.394 & 1.897 & .058 \\




Previous Awareness  (Partial) 
  & 0.791 & 0.320 & 2.471 & <.05
  & 0.193 & 0.402 & 0.481 & .631 \\

Previous Awareness (Full) 
  & 0.253 & 0.324 & 0.780 & .435
  & -1.227 & 0.417 & -2.941 & <.01 \\

\midrule

\textbf{Random Effects} 
  & \textbf{Var} & \textbf{SD} &  &  
  & \textbf{Var} & \textbf{SD} &  &  \\ 
\midrule
Participant 
  & 2.392 & 1.547 &  &  
  & 1.699 & 1.304 &  &  \\
Item 
  & 0.000 & 0.000 &  &  
  & 0.030  & 0.174 &  &  \\
\midrule

\textbf{Model Fit} 
   & \textbf{logLik} & \textbf{AIC} &  &
   & \textbf{logLik} & \textbf{AIC} &  & \\ 
\midrule
   & -501.45 & 1046.90 &  &
   & -292.61 & 625.21 &  &  \\
\bottomrule
\end{tabular}
\end{table}

\begin{table}[h!]
\centering
\caption{This table presents the results of the CLMM analysis for Model 3.}
\label{tab:A3}
\Description{This table presents the results of the cumulative link mixed model (CLMM) analysis for Model 3, which investigates how the perceived severity of the current and previous failures influences trust and perceived intelligence ratings. The table reports estimated coefficients, standard errors, z-values, and p-values for the main predictors, including current and previous failure severity and their interaction. Covariates such as previous rating, failure number, and current and previous awareness levels are also included. Random effects for participants and items are shown, along with model fit statistics (log-likelihood and AIC) for both trust and perceived intelligence outcomes.}

\renewcommand{\arraystretch}{1.2}
\setlength{\tabcolsep}{6pt}

\begin{tabular}{
    >{\centering\arraybackslash}m{4.2cm} |
    >{\centering\arraybackslash}m{0.93cm} 
    >{\centering\arraybackslash}m{0.8cm} 
    >{\centering\arraybackslash}m{0.8cm} 
    >{\centering\arraybackslash}m{0.8cm} |
    >{\centering\arraybackslash}m{0.93cm} 
    >{\centering\arraybackslash}m{0.8cm} 
    >{\centering\arraybackslash}m{0.8cm} 
    >{\centering\arraybackslash}m{0.8cm} 
}
\toprule
\textbf{Model 3} 
  & \multicolumn{4}{c|}{\textbf{Trust}} 
  & \multicolumn{4}{c}{\textbf{Perceived Intelligence}} \\ 
\cmidrule(lr){2-9}
\textbf{Fixed Effects} 
  & \textbf{Est} & \textbf{SE} & \textbf{z} & \textbf{p}
  & \textbf{Est} & \textbf{SE} & \textbf{z} & \textbf{p} \\ 
\midrule

\multicolumn{9}{l}{\textbf{Main Predictors}} \\
Severity Current Failure (High) 
  & -1.267 & 0.788 & -1.609 & .108
  & -1.514 & 0.723 & -2.093 & <.05 \\
Severity Previous Failure (High) 
  & 1.728 & 0.791 & 2.184 & <.05
  & 0.932 & 0.741 & 1.258 & .209 \\
Current (High) × Previous Failure Severity (High) 
  & -1.119 & 0.871 & -1.284 & .199
  & -0.855 & 0.836 & -1.023 & .306 \\
\midrule

\multicolumn{9}{l}{\textbf{Covariate/Control Factors}} \\
Previous Rating 
  & 0.738 & 0.153 & 4.812 & <.001
  & 1.052 & 0.208 & 5.057 & <.001 \\

Failure Number (Third) 
  & -0.024 & 0.192 & -0.124 & 0.901
  & -0.126 & 0.240 & -0.525 & .599 \\


Current Awareness (Partial)
  & 1.172 & 0.306 & 3.833 & <.001
  & 0.841 & 0.363 & 2.320 & <.05 \\
Current Awareness (Full) 
  & 1.972 & 0.324 & 6.085 & <.001
  & 1.524 & 0.396 & 3.846 & <.001 \\


Previous Awareness  (Partial) 
  & 0.871 & 0.327 & 2.710 & <.01
  & 0.210 & 0.395 & 0.531 & .595 \\
Previous Awareness (Full) 
  & 0.407 & 0.327 & 1.246 & .213
  & -1.089 & 0.406 & -2.685 & <.01 \\

\midrule

\textbf{Random Effects} 
  & \textbf{Var} & \textbf{SD} &  &  
  & \textbf{Var} & \textbf{SD} &  &  \\ 
\midrule
Participant 
  & 2.467 & 1.571 &  &  
  & 1.548 & 1.244 &  &  \\
Item 
  & 0.000 & 0.000 &  &  
  & 0.028 & 0.169 &  &  \\
\midrule

\textbf{Model Fit} 
   & \textbf{logLik} & \textbf{AIC} &  &
   & \textbf{logLik} & \textbf{AIC} &  & \\ 
\midrule
   & -518.75 & 1069.51 &  &
   & -304.18 & 638.36 &  &  \\
\bottomrule
\end{tabular}
\end{table}

\end{document}